\documentclass[english,tightenlines,eqsecnum,amsmath,amssymb,aps,prd,superscriptaddress,nofootinbib,showpacs,floats]{article}

\usepackage{epsfig}
\usepackage{babel}
\usepackage{authblk}
\usepackage{amsmath,amsfonts,amssymb,multicol}
\usepackage[usenames]{xcolor}
\definecolor{AHZ}{rgb}{0.0,0.0,0.9}
\definecolor{AHZ1}{rgb}{1,0.0,0.1}
\usepackage[colorlinks,linkcolor=AHZ,citecolor=AHZ1,bookmarks]{hyperref}
\long\def\/*#1*/{}
\usepackage{color}

{\definecolor{RED}{rgb}{1,0,0}
\definecolor{GREEN}{rgb}{0,1,0}
\definecolor{BLUE}{rgb}{0,0,1}
\usepackage{graphicx,caption,subcaption}

\begin{document}

\title{Late-time cosmological evolution of a general class of $f({\sf R},{\sf T})$ gravity with minimal curvature-matter coupling}
\author[1]{Hamid Shabani,\thanks{h.shabani@phys.usb.ac.ir}}
\author[2]{Amir Hadi Ziaie,\thanks{ah.ziaie@gmail.com}}

\affil[1]{Physics Department, Faculty of Sciences, University of Sistan and Baluchestan, Zahedan, Iran}
\affil[2]{Department of Physics, Kahnooj Branch, Islamic Azad University, Kerman, Iran}
\date{\today}
%
\maketitle
\begin{abstract}
\noindent
In this work, we study the late time cosmological solutions of $f({\sf R},{\sf T})=g({\sf R})+h(-{\sf T})$ models 
assuming that the conservation of energy-momentum tensor ({\sf EMT})
is violated. We perform our analysis through constructing an autonomous dynamical system for
equations of motion. We study the stability properties of solutions via
considering linear perturbations about the related equilibrium points. Moreover, we parameterize the
Lagrangian by introducing the parameters $m(r)$ and $n(s)$. These parameters
which are constructed out of the functions $g({\sf R})$ and $h(-{\sf T})$ play the main role in finding the
late time behavior of the solutions. We find that there exist, in general, three classes of solutions; all models with
$n>0$ include a proper transition from a prolonged matter era to a de Sitter solution.
Models with $-0.5<n<0$ and $n'>1$, for at least a root of equation $n(s)=s-1$, include
an unphysical dark energy solution preceding an improper matter era. Finally, for $n<-1/2$
there is a transient accelerated expansion era with $-1/2<w^{\rm{(eff)}}<-1/3$ before a de Sitter phase.
For all cases, in order to have a long enough matter dominated epoch, the condition $m'\to 0^{+}$ for $r\lessapprox-1$ must hold.
We also show that models with power law dependency i.e., $f({\sf R},{\sf T})={\sf R}^{\beta}+(-{\sf T})^{\alpha}$ can be observationally motivating for $m\to0^{+}$ and $-0.024<\alpha<0.02$ and therefore could provide a suitable setting for later investigations.
\end{abstract}
\section{Introduction}\label{intro}
One of the recent major challenges in modern cosmology is the late time behavior of the Universe. Using distant type Ia supernovae as standard candles, it has been discovered that the Universe is currently experiencing an accelerated expansion~\cite{supno3,SDSS2,WMAP5,Planck}. Such a discovery led to the widespread acceptance of the idea that the Universe is dominated by a mysterious substance (with negative pressure) named \lq\lq{}dark energy\rq\rq{} ({\sf DM}) that drives this accelerating expansion. It is a general belief that the concordance model~\cite{LCDM} which includes simply cosmological constant within the Einstein\rq{}s field equation is also well consistent with the most observational data.
Therefore, with this assumption for which all analyses of the recent data have been performed, the equation of state parameter ({\sf EoS}) of {\sf DE} is found to be $ w^{\rm{({\sf DE})}}= −1.006\pm0.045$~\cite{Planck}. However, there are some substantial difficulties within the cosmological constant description of late time behavior of the Universe. The well-known cosmological constant problem states that the concordance model suffers from a fine-tuning issue related to its energy scale, if it is attributed to vacuum energy~\cite{Cos.pro1,Cos.pro3}. Indeed, from
a theoretical viewpoint, the necessity of explaining {\sf DE}, as well as the second dominant component of the Universe, i.e., dark matter ({\sf DM}) \cite{darkmatt}, raises the fundamental question of whether the general theory of relativity ({\sf GR}) could provide
a befitting quantitative description of the Universe on all scales.
In general, there are two basic approaches that could provide a setting for theoretical explanation
of the accelerated expansion of the Universe. The first procedure consists of modifying the matter content of the
Universe by introducing a {\sf DE} sector, starting either with a canonical scalar
field, a phantom field, or the combination of both fields in a unified model and proceeding afterwards
to more complicated scenarios; see \cite{DMsector} and references therein for more details and reviews.
The second approach is to modify the gravitational sector itself (see e.g. \cite{gravitsect,fR,grsect100,grsect101}), motivated by this fundamental assumption that at large astrophysical and cosmological scales, standard {\sf GR} may not describe correctly the dynamical evolution of the Universe.
To deal with this issue, several efforts have been made, among which gravity theories extending {\sf GR} have  attracted a great deal of interest over the past decades. In the context of modified gravity theories ({\sf MGT}), a geometric description for {\sf DM} can be provided and the accelerated expansion can be achieved at late times, thus, the problem of cosmological constant may be resolved (for a detailed dynamical system analysis of several cosmological models in the context of {\sf MGT} see \cite{saridynsys}.) Various methods have been proposed so far, in order to modify the gravitational action (cf. \cite{CLFEPASK} for a consummate review), giving rise to different classes of {\sf MGT}. From a historical perspective, in going beyond the {\sf GR} theory, the first step has been to generalize the geometric part of the Einstein-Hilbert action. One of the simplest forms with the aim of modifying the dynamical behavior of gravity and consequently reaching cosmological scenarios different to {\sf GR}, is to consider functions of the Ricci scalar ${\sf R}$, dubbed $f({\sf R})$ theories~\cite{fR}. Recently, this picture has been extended to incorporate the trace of energy momentum tensor ({\sf EMT}) within the gravitational sector of the action. Such a class of {\sf MGT} known as \lq\lq{}$f({\sf R},{\sf T})$ gravity theories\rq\rq{} which allow for matter-geometry coupling are significant as they can provide, from a fundamental theoretical perspective, a complete theoretical description for the late time acceleration of the Universe, without resorting to the existence of {\sf DE} \cite{fRT1}. These models can also offer some alternative explanations for the nature of {\sf DM} \cite{Harko14,Zaregonbadi} and the problem of flattening of the galaxy rotation curves \cite{bertorotationc}. Since the advent of this idea, different features of $f({\sf R},{\sf T})$ gravity have been investigated and in particular many efforts have been devoted to studying the cosmological aspects of this theory~\cite{fRT1,fRT2,fRT3,fRT4,fRT5,fRT6,fRT7,fRT8,fRT9,fRT10,fRT11,fRT12,fRT13,fRT14,fRT15} and very recently, quantum cosmology of $f({\sf R},{\sf T})$ gravity has been proposed in~\cite{fRT16} where the authors have investigated the evolution of the wave function of early Universe by virtue of the Wheeler-de Witt equation. In~\cite{fRT2,fRT4}, the background evolution of the Universe has been studied under the assumption that the {\sf EMT} is conserved. Taking into account the conservation of {\sf EMT} as a basic constraint, the authors have shown that $f({\sf R},{\sf T})$ models with a minimal coupling between the geometrical and matter sectors yield the late time accelerated expansion of the Universe. However, most of these models lead to the present value for the {\sf EoS} parameter as $w^{\rm{({\sf DE})}}=-1/2$, which is not observationally acceptable. Therefore, it is reasonable to seek alternative ways in order to cure this problem.

Very recently, an interesting result has been reported in~\cite{josset} where the authors have found that  in the context of unimodular gravity, violation of {\sf EMT} conservation can lead to the emergence of an effective cosmological constant. Regardless of how the conservation of {\sf EMT} can be violated,\footnote{For example, in the context of classical {\sf GR}, the violation of {\sf EMT} conservation could provide a setting to deal with the entropy problem of standard cosmology, see e.g. \cite{violationcGR} and references therein. Such a violation may also occur by modifying quantum mechanical models and also some proposals based on the causal set approach to quantum gravity \cite{josset}.} it is a novel idea to reconsider previous unsuccessful models under this assumption. It should be emphasized that, though the violation of {\sf EMT} conservation may not always lead to a physically reasonable accelerated expansion scenario at late time, it can be a novel task to investigate the effects of violation of {\sf EMT} conservation on the late time behavior of a gravitation theory. Motivated by this idea, we study minimally coupled models of the form $f({\sf R},{\sf T})={\sf R}+\Lambda({\sf T})$ as presented in~\cite{fRT12}, ignoring the usual relation for {\sf EMT} conservation. These types of models can account for, at least, modification of the Einstein-Hilbert action by putting aside the assumption that the matter Lagrangian plays a subordinate and passive role only. In the present paper we extend the previous work to include models with arbitrary functions of the Ricci curvature scalar. We shall see that the violation of {\sf EMT} conservation could give rise to acceptable accelerated expansion at late times, contrary to the situations where the conservation of {\sf EMT} is respected. Note that in $f({\sf R})$ gravity, applying the Bianchi identity to the geometrical parts of the field equation leads to a null result and, thus, the condition on {\sf EMT} conservation cannot be relaxed. From this point of view, $f({\sf R},{\sf T})$ gravity has more chance for designing viable cosmological models.

Up to now, the dynamical system approach has been widely used to describe cosmological aspects of different gravity theories~\cite{saridynsys}. Using this powerful method, one can find cosmological solutions and investigate the nature of the fixed points and their stability properties. However, this approach has not been completely utilized in the framework of $\sf{f(R,T)}$ gravity. Nevertheless, dynamical system techniques have been exploited to investigate cosmological behavior of models of the form $\sf{f(R,T)=R+C\sqrt{T}}$~\cite{fRT2,fRT4}, models that deal with the Einstein static Universe~\cite{fRT7} and late time solutions for $\sf{f(R,T)=R+h(T)}$ gravity models~\cite{Baffou14,fRT12}. The present work could be among the first studies which deal with the dynamical system approach in the context of $\sf{f(R,T)=g(R)+h(T)}$ models with the aim to describe the late time cosmological acceleration. We would like to add that there are different works in the literature where the authors have considered cosmological acceleration in ${\sf f(R,T)}$ gravity, however, they have employed algebraic manipulations or have obtained cosmological solutions by applying indirect procedures such as reconstructing ${\sf f(R,T)}$ models for a presumed cosmological behavior~\cite{Jamil12,Houndjo12,Singh14,Sharif142,Sharif141,Rudra15,Baffou15,Moraes161}. A significant trait of our work is a parametrization of the variables which is used in the dynamical field equations. We shall show that in $f(\sf{R},\sf{T})=g(\sf{R})+h(\sf{T})$ gravity models, one can recast the functions constructed out of curvature and the trace of {\sf EMT}, i.e.,  $g(\sf{R})$ and $h(\sf{T})$ (at least for well-defined ones) into the four dimensionless parameters $(m,r)$ and $(n,s)$, respectively. More exactly, one can rewrite the expressions given for $g(\sf{R})$ and $h(\sf{T})$ in terms of equivalent dimensionless $m(r)$ and $n(s)$ functions.  Therefore, when these parameters appear in the dynamical equations, the critical points and their stability properties could determine consistent ranges of these parameters and thus the validity of the underlying model. Most studies in the literature either deal with particular models from the beginning of the problem or present a general form of the field equations and finally employ them for some specific types of $g(\sf{R})$ and $h(\sf{T})$ functions. For instance, authors of~\cite{SinghP14,Moraes15,Yadav15,Moraes162} have discussed models with $h(\sf{T})=\alpha \sf{T}$. The case of a power law form, $h(\sf{T})=\sf{T}^\alpha$, has been considered in~\cite{Sun16,Baffou171} and investigation of some non-trivial forms has been carried out in~\cite{Moraes161,Moraes163, Baffou172}. Briefly speaking, we use a formalism which works at least for a wide well-defined kinds of Lagrangians. This formalism was introduced for the first time in~\cite{amend}.

This paper is arranged as follows: In Sect.~\ref{Fieldequations}, the field equations for $f({\sf R},{\sf T})$ gravity together with essential variables will be introduced. In this section we also discuss the violation of {\sf EMT} conservation. In Sect.~\ref{gen} we construct an autonomous system of equations and consider the corresponding fixed point solutions. Moreover,  we study the cosmological behavior of the minimal $f({\sf R},{\sf T})$ models by relaxing the conservation of {\sf EMT}. In Sect.~\ref{pure}, we explore models with power law form for the Lagrangian. Finally in Sect.~\ref{con}, we summarize our results.
\section{Field equations of $f({\sf R},{\sf T})$ gravity}\label{Fieldequations}
In the present section, we review the field equations of $f({\sf R},{\sf T})$ gravity and introduce necessary
variables which shall be used in the rest of paper. Since we are interested in studying the late time solutions of a
general class of $f({\sf R},{\sf T})=g({\sf R})+h({\sf T})$ models, we assume only the pressure-less fluid as the matter content. We call this particular model the \lq\lq{}minimal $f({\sf R},{\sf T})$ model\rq\rq{} for, only the minimal coupling between geometrical and matter sectors is taken into account. $f({\sf R},{\sf T})$ gravity theories have been initially introduced by the following action~\cite{fRT1}
\begin{align}\label{action}
S=\int \sqrt{-g} d^{4} x \left[\frac{1}{2\kappa^{2}} f\Big{(}{\sf R}, {\sf T}\Big{)}+{\sf L}^{\rm{(m)}}\right],
\end{align}
where ${\sf R}$, ${\sf T}\equiv g^{\mu \nu} {\sf T}_{\mu \nu}$, ${\sf L}^{\rm{(m)}}$ are defined as the Ricci
curvature scalar, the trace of energy momentum tensor and the Lagrangian of
pressure-less matter (here, pressure-less fluid), respectively. $g$ is the determinant of the metric,
$\kappa^{2}\equiv8 \pi G$ and we set $c=1$. The {\sf EMT} of matter, ${\sf T}_{\mu \nu}$ is defined as the Euler-Lagrange expression of the matter Lagrangian, i.e.,
\begin{align}\label{Euler-Lagrange}
{\sf T}_{\mu \nu}\equiv-\frac{2}{\sqrt{-g}}
\frac{\delta\left[\sqrt{-g}{\sf L}^{\rm{(m)}}\right]}{\delta g^{\mu \nu}}.
\end{align}
A point here which begs a few elucidation is the inclusion of trace of {\sf EMT} within the curvature part of the action. This might, at first sight, be seen a little ambiguous as one usually expects that the matter Lagrangian or a function of it must be coupled to an arbitrary function of the Ricci scalar. In this respect, non-trivial gravitational Lagrangians have been proposed where the gravitational sector directly interacts with matter sector through a non-minimal coupling process, the so-called $f({\sf R},{\sf L}_m)$ gravity \cite{berto-nonmini-cappo}. Various choices for matter Lagrangian would lead to different gravitational theories and thus have various cosmological scenarios as the outcome. In this regard, one possible way to take into account the footprint of matter within the gravitational Lagrangian is to include the trace of {\sf EMT} within the gravitational sector of the action.\footnote{ The dependence of the gravitational action on the trace
of {\sf EMT} may be due to the presence of quantum effects (conformal anomaly), or of some exotic imperfect quantum fluids; see e.g, \cite{Harko14} for more details.} This can be regarded as a possible way of defining the gravitational Lagrangian (\ref{action}) where one assumes that the contributions due to matter that appear within the argument of the function $f$ solely correspond to the {\sf EMT} constructed out of the material part of the Lagrangian, as defined via (\ref{Euler-Lagrange}). We then might intuitively imagine that the matter existing within the spacetime fabric (as descried by the Lagrangian ${\sf L}_m$) creates curvature. The matter content then interacts minimally with spacetime curvature through the metric and curvature mutually has an effect on the matter distribution within the spacetime. However, the {\sf EMT} trace is taken a {\it priori} as the only agent for direct interaction between matter and curvature. The {\it a priori} contribution owing to matter in an uncommon interaction with spacetime curvature may also have some connections with known subjects such as a geometrical description of the curvature induced due to matter distribution, describing physical forces geometrically, and a geometrical origin for the matter content of the Universe; see, e.g., \cite{farhoudipps} and the references therein. A specific class of $f({\sf R},{\sf L}_m)$ gravity theories can be constructed assuming that the matter Lagrangian is a function of energy density of matter $\rho$ only, so that ${\sf L}_m={\sf L}_{m}(\rho)$. The {\sf EMT} then reads \cite{Kharko}
\begin{align}\label{emtlmrho}
{\sf T}_{\mu\nu}=-\rho\frac{d{\sf L}_m}{d\rho}{\sf U}_\mu{\sf U}_\nu+\left[{\sf L}_m-\rho\frac{d{\sf L}_m}{d\rho}\right]g_{\mu\nu},
\end{align}
where ${\sf U}_\mu$ being the four-vector velocity of the fluid which satisfies the condition ${\sf U}_\mu{\sf U}^\mu=-1$.
Considering, for instance, the choice of ${\sf L}_m=-\rho$ for matter Lagrangian \cite{Brown-HAwking,Brown-HAwking1}, we get the {\sf EMT} for a pressure-less fluid as
\begin{align}\label{pffluid}
{\sf T}_{\mu\nu}=\rho{\sf U}_\mu{\sf U}_\nu.
\end{align}
The {\sf EMT} trace for such a fluid (which is our case study) is found as ${\sf T}=-\rho$, and this could justify the way of the appearance of the {\sf EMT} trace within the gravitational sector of the action.

Varying action (\ref{action}) with respect to the metric field leads to the following field equation~\cite{fRT1}
\begin{align}\label{fRT field equations}
&F({\sf R},{\sf T}) {\sf R}_{\mu \nu}-\frac{1}{2} f({\sf R},{\sf T}) g_{\mu \nu}+\Big{(} g_{\mu \nu}
\square -\triangledown_{\mu} \triangledown_{\nu}\Big{)}F({\sf R},{\sf T})=\nonumber\\
&\Big{(}\kappa^2-{\mathcal F}({\sf R},{\sf T})\Big{)}{\sf T}_{\mu \nu}-\mathcal {F}({\sf R},{\sf T})\mathbf
{\Theta_{\mu \nu}},
\end{align}
where
\begin{align}\label{theta}
\mathbf{\Theta_{\mu \nu}}\equiv g^{\alpha \beta}\frac{\delta
{\sf T}_{\alpha \beta}}{\delta g^{\mu \nu}},
\end{align}
and for the sake of convenience, we have defined the following functions for derivatives with respect
to the trace ${\sf T}$ and the Ricci curvature scalar ${\sf R}$, as
\begin{align}\label{f definitions1}
{\mathcal F}({\sf R},{\sf T}) \equiv \frac{\partial f({\sf R},{\sf T})}{\partial {\sf T}}~~~~~
~~~~~\mbox{and}~~~~~~~~~~
F({\sf R},{\sf T}) \equiv \frac{\partial f({\sf R},{\sf T})}{\partial {\sf R}}.
\end{align}
We assume that the spacetime geometry is described by the spatially flat Friedmann--Lema\^{\i}tre--Robertson--Walker ({\sf FLRW}) metric, which in spherical coordinates is given by
\begin{align}\label{metricFRW}
ds^{2}=-dt^{2}+a^{2}(t) \Big{(}dr^{2}+r^{2}d\Omega^2\Big{)},
\end{align}
where $a$ denotes the scale factor of the Universe and $d\Omega^2$ being the line element on a unit two-sphere. The field equation (\ref{fRT field equations}) then gives rise to
\begin{align}\label{first}
&3H^{2}F({\sf R},{\sf T})+\frac{1}{2} \Big{(}f({\sf R},{\sf T})-F({\sf R},{\sf T}){\sf R}\Big{)}+3\dot{F}({\sf R},{\sf T})H=\nonumber\\
&\Big{(}\kappa^2 +{\mathcal F} ({\sf R},{\sf T})\Big{)}\rho,
\end{align}
as the modified Friedmann equation,
\begin{align}\label{second}
&2F({\sf R},{\sf T}) \dot{H}+\ddot{F} ({\sf R},{\sf T})-\dot{F} ({\sf R},{\sf T}) H=-\Big{(}\kappa^2+{\mathcal F} ({\sf R},{\sf T})\Big{)}\rho,
\end{align}
as the modified Raychaudhuri equation and $H$ indicates the Hubble parameter. 

In the case of $f({\sf R},{\sf T})$ gravity one may define a {\sf DE} component
which is responsible for the accelerated expansion at the late times. To this aim, the Friedmann equation, (\ref{first}), and the Raychaudhuri equation (\ref{second}) can be redefined as \cite{fRT4}\footnote{This representation has been already used in the literature 
for $f({\sf R})$ gravity; for more details we consult the reader to~\cite{Hwang,Amend,Fay,Elizalde,Oikonomou,Mukherjee,Bamba,Cosmai,Battye}.}
\begin{align}\label{redefinition1}
3H^2F({\sf R},{\sf T})=8\pi G \left(\rho+\rho^{\textrm{({\sf DE})}}\right),
\end{align}
and
\begin{align}\label{redefinition2}
-2\dot{H}F({\sf R},{\sf T})=8\pi G \Big{(}\rho+\rho^{\textrm{({\sf DE})}}+p^{\textrm{({\sf DE})}}\Big{)},
\end{align}
where in equations (\ref{redefinition1}) and  (\ref{redefinition2}) the first term
is related to the pressure-less fluid and $\rho^{\textrm{({\sf DE})}}$ and $p^{\textrm{({\sf DE})}}$
denote the density and pressure of {\sf DE} component, respectively, which are defined as
\begin{align}\label{DEd}
8\pi G\rho^{\textrm{({\sf DE})}}\equiv{\mathcal F}({\sf R},{\sf T})\rho^{({\sf m})}-3\dot{F}({\sf R},{\sf T})H-\frac{1}{2}\Big{(}f({\sf R},{\sf T})-F({\sf R},{\sf T}){\sf R}\Big{)},
\end{align}
and
\begin{align}\label{DEp}
8\pi Gp^{\textrm{({\sf DE})}}\equiv\ddot{F}({\sf R},{\sf T})+2\dot{F}({\sf R},{\sf T})H+\frac{1}{2}\Big{(}f({\sf R},{\sf T})-F({\sf R},{\sf T}){\sf R}\Big{)}.
\end{align}
To see the consistency of the above definitions, one can substitute the Lagrangian for {\sf GR}, i.e., $f({\sf R},{\sf T})={\sf R}$ into equations (\ref{DEd}) and (\ref{DEp}). This leaves us with the usual Friedman equations instead of those given in  (\ref{redefinition1}) and  (\ref{redefinition2}), and also $\rho^{\textrm{({\sf DE})}}=0, p^{\textrm{({\sf DE})}}=0$ in the case of {\sf GR}. In fact, these definitions introduce the {\sf DE} component as a real fluid\footnote{As we shall see, interactions of these two components needs to define an effective behavior and therefore effective pressure and density.}.

  In this case, the {\sf EoS} parameter for {\sf DE} is defined as $w^{\textrm{({\sf DE})}}\equiv p^{\textrm{({\sf DE})}} / \rho^{\textrm{({\sf DE})}}$, as usual. 
One can check that definitions (\ref{DEd}) and (\ref{DEp}) guarantee the 
continuity equation for {\sf DE} component, i.e.,
\begin{align}\label{DEw}
\dot{\rho}^{\textrm{({\sf DE})}}+3H\left(1+w^{\textrm{({\sf DE})}}\right)\rho^{\textrm{({\sf DE})}}=0.
\end{align}
In the context of modified gravity theories (specially in $f({\sf R})$ gravity), an effective {\sf EoS} parameter
would be defined very similar to the {\sf GR} case. In {\sf GR}, the {\sf EoS} parameter for a perfect fluid is defined as
$w=p/\rho$. Using this definition in modified gravity, one can consider an effective behavior of fluids and correspondingly define an effective {\sf EoS} as $w^{\textrm{({\sf eff})}}\equiv -1-2 \dot{H}/3 H^2$. Performing
some straightforward calculations we conclude that
\begin{align}\label{weffd}
w^{\textrm{({\sf eff})}}=\Omega^{\textrm{({\sf DE})}} w^{\textrm{({\sf DE})}},
\end{align}
where we have defined a density parameter for {\sf DE} component as 
$\Omega^{\textrm{({\sf DE})}}\equiv 8\pi G\rho^{\textrm{({\sf DE})}}/3H^{2}F$. The
effective {\sf EoS} includes information about gravitational interactions for 
different components (pressure-less and {\sf DE} in our case). More precisely, one can define 
an effective density and pressure using $3H^{2}=8\pi G \rho^{\textrm{({\sf eff})}}$ and $-2\dot{H}=8\pi G (\rho^{\textrm{({\sf eff})}}+p^{\textrm{({\sf eff})}})$, thereby, it is possible to describe the cosmological consequences of the theory with the help of the effective behavior of the pressure-less and {\sf DE} components.\footnote{Redefinition of the Friedman equations of $f({\sf R},{\sf T})$ gravity in terms of an conserved effective fluid could lead to interesting results. This approach have let us to discover some novel features of $f({\sf R},{\sf T})$ gravity~\cite{Shabani6}.}

Next, we proceed to consider a Lagrangian with minimal coupling between the Ricci scalar and the trace of {\sf EMT}, i.e.,\footnote{We do not
include a coupling constant, since, such a coefficient could be absorbed into the dynamical
system variables.}
\begin{align}\label{minimal}
f({\sf R},{\sf T})=g({\sf R})+h(-{\sf T}).
\end{align}
Because of our metric signature, we have ${\sf T}=-\rho$ for the pressure-less matter, therefore,
to avoid ambiguity due to the negative sign, we hereafter study $h(-{\sf T})$ functions.
For this class of $f({\sf R},{\sf T})$ models we obtain $\mathcal{F}({\sf R}, {\sf T})=-h'(-{\sf T})$ and
$F({\sf R}, {\sf T})=g'({\sf R})$, hence the field equations
(\ref{first}) and (\ref{second}) yield
\begin{align}\label{eom1}
&1+\frac{g}{6H^{2} g'} +\frac{h}{6 H^{2} g'}-\frac{{\sf R}}{6 H^{2}} + \frac{\dot{g'}}
{H g'}=\frac{\kappa^2 \rho}{3H^{2} g'}-\frac{h' \rho}{3H^{2} g'},
\end{align}
and
\begin{align}\label{eom2}
&2\frac{\dot{H}}{H^{2}}+\frac{\ddot{g'}}{H^{2} g'} -\frac{\dot{g'}}{H g'}=
-\frac{\kappa^2 \rho}{H^{2} g'}+\frac{h' \rho}{H^{2} g'}.
\end{align}
Let us now define a few variables and parameters which are useful for recasting the field equations
(\ref{eom1}) and (\ref{eom2}) into a closed dynamical system. These variables are defined as
\begin{align}
&x_{1}\equiv-\frac{\dot{g}({\sf R})}{H g'({\sf R})},\label{varx1}\\
&x_{2}\equiv-\frac{g({\sf R})}{6 H^{2} g'({\sf R})},\label{varx2}\\
&x_{3}\equiv \frac{{\sf R}}{6 H^{2}}=\frac{\dot{H}}{H^{2}}+2,\label{varx3}\\
&x_{4}\equiv -\frac{h(-{\sf T})}{6H^{2} g'({\sf R})},\label{varx4}\\
&x_{5}\equiv \frac{{\sf T} h'(-{\sf T})}{3H^{2} g'({\sf R})},\label{varx5}\\
&\Omega^{{\sf dust}}\equiv\frac{\kappa^{2} \rho}{3 H^{2} g'({\sf R})},\label{omega mat}
\end{align}
where we have used the expression ${\sf R}=6(\dot{H}+2H^2)$ for the Ricci scalar within the definition ($\ref{varx3}$). Using definition (\ref{varx3}) along with the mentioned definition of {\sf EoS}, we obtain
\begin{align}\label{weff2}
w^{\textrm{({\sf eff})}}=\frac{1}{3}\left(1-2x_{3}\right).
\end{align}
Also, using the definitions (\ref{DEd}), (\ref{varx1})-(\ref{omega mat}) and the density parameter of {\sf DE}
which is defined in (\ref{weffd}), we get 
\begin{align}\label{weff2}
&\Omega^{({\sf DE})}=x_{1}+x_{2}+x_{3}+x_{4}+x_{5},\\
&\Omega^{{\sf dust}}+\Omega^{({\sf eff})}=1.
\end{align}
Furthermore, we use the following definitions
\begin{align}
&m \equiv \frac{{\sf R} g''({\sf R})}{g'({\sf R})},\label{m}\\
&r \equiv -\frac{{\sf R} g'({\sf R})}{g({\sf R})}=\frac{x_{3}}{x_{2}}\label{r},\\
&n \equiv -\frac{{\sf T} h''(-{\sf T})}{h(-{\sf T})}\label{n},\\
&s \equiv -\frac{{\sf T}h'(-{\sf T})}{h(-{\sf T})}=\frac{x_{5}}{2x_{4}},\label{s}
\end{align}
where all primes denote differentiation with respect to the argument. Expressions (\ref{m})-(\ref{s}) show that eliminating ${\sf R}$ from (\ref{m}) and (\ref{r}) leads to the function $m=m(r)$ and also eliminating ${\sf T}$ from (\ref{n}) and (\ref{s}) gives
the function $n=n(s)$. As we mentioned before, parameters $r$ and $m$ have originally been used in~\cite{amend},
in order to describe cosmological solutions of $f({\sf R})$ gravity models. Here, we extend their method
to the case of minimal $f({\sf R},{\sf T})$ model. Models with $m=m(r)$ and $n=n(s)$, instead of $g({\sf R})$ and $h(-{\sf T})$, can be very suitable for dynamical system analysis.

In~\cite{fRT2,fRT4}, the authors have comprehensively considered cosmological solutions
of $f({\sf R},{\sf T})$ gravity when the conservation of {\sf EMT} is respected. To see the consequences of
{\sf EMT} conservation, one can apply the Bianchi identity to the field equation (\ref{fRT field equations}). Then imposing the usual form for the {\sf EMT} conservation of matter current i.e., $\nabla_{\alpha}{\sf T}^{\alpha\beta}=0$ which leads to $\dot{\rho}+3H\rho=0$, and applying the Bianchi identity to the field equation (\ref{fRT field equations}), we arrive at the following simple constraint:
\begin{align}\label{constraint0}
\frac{1}{2}\mathcal {F}\nabla_{\mu}{\sf T}
+{\sf T}_{\mu \nu}\nabla^{\mu}\mathcal {F}=0,
\end{align}
whereby substituting for the metric components gives
\begin{align}\label{constraintt}
\dot{\mathcal{F}}({\sf R},{\sf T})=\frac{3}{2}H\mathcal{F}({\sf R},{\sf T}).
\end{align}
As is seen, equation (\ref{constraintt}) places a constraint on the functionality of $f({\sf R},{\sf T})$, which in the case of the minimal coupling directly leads to a constraint on the trace dependent sector (as we shall see below), since, as is well known, in pure $g({\sf R})$ gravity the Bianchi identity is automatically satisfied, therefore leading to the conservation of {\sf EMT}. Substituting the Lagrangian (\ref{minimal}) in constraint~(\ref{constraintt}) gives
\begin{align}\label{constraint2}
\dot{h}'(-{\sf T})=\frac{3}{2}Hh'(-{\sf T}),
\end{align}
whereby after a straightforward algebra we get
\begin{align}\label{specific}
h(-{\sf T})=C_{1}\sqrt{-{\sf T}}+C_{2},
\end{align}
where $C_{1}$ and $C_{2}$ are constants of integration. Therefore, constraint~(\ref{constraintt}) 
determines the form of $f({\sf R},{\sf T})$ function and in the minimally coupled case this constraint 
gives the relation~(\ref{specific}). We note that this solution with the use of definition (\ref{n}) corresponds to $n=-1/2$. The  behavior of these types of models has been fully studied in~\cite{fRT2,fRT4} and further 
issues discussed in~\cite{fRT12} using a dynamical system approach. It is shown that
despite the appearance of accelerated expansion at late times, the corresponding dark
energy cannot be suitably matched with the observational data. In these models a {\sf DE} with $w^{\rm{({\sf DE})}}_{0}=-1/2$ is responsible for the accelerated expansion of the Universe. Hence, it may be a good idea to explore the cosmological behavior of the model under violation of {\sf EMT} conservation. To this aim, we rewrite the field equation (\ref{first}) for specific models given by the choice (\ref{minimal}) for $f({\sf R},{\sf T})$ function, as follows:
\begin{align}\label{fieldmini}
 {\sf R}_{\mu \nu}-\frac{1}{2}\Big{(} {\sf R}+h\Big{)} g_{\mu \nu}=\Big{(}\chi^{2}-h'\Big{)} {\sf T}_{\mu \nu}.
\end{align}
Then, we apply the Bianchi identity to the field equation (\ref{fieldmini}) whence we obtain the following covariant conservation equation
\begin{align}\label{relation}
&(\kappa^{2} -h')\nabla^{\mu}{\sf T}_{\mu \nu}-\frac{1}{2}h'\nabla_{\mu}{\sf T}
-{\sf T}_{\mu \nu}\nabla^{\mu}h'=0,
\end{align}
where the argument of $h'(- {\sf T})$ has been dropped for abbreviation. Note that equation  (\ref{relation}) is more general than equation (\ref{constraint0}), which holds when we do not assume the conservation of matter {\sf EMT}, i.e., $\nabla_{\alpha}{\sf T}^{\alpha\beta}=0$. Substituting for the components of {\sf EMT} of a dust into equation (\ref{relation}), we obtain
\begin{align}\label{relation-3}
\Big{(}\kappa^{2} -\frac{3}{2}h'+h''{\sf T}\Big{)}\dot{{\sf T}}+3H{\sf T}\Big{(}\kappa^{2} -h'\Big{)}=0,
\end{align}
where we have used $\rho=-{\sf T}$. Comparing equation (\ref{constraint2}) and (\ref{relation-3}) shows that, in the case of relaxing the conservation of {\sf EMT}, a more complicated constraint would determine the form of $h(-{\sf T})$ function. Once the function $h(-{\sf T})$ is determined, the dependency of $-{\sf T}$ and consequently $\rho$ on the scale factor can be calculated. On the other hand, in two situations equation (\ref{relation-3}) could simply lead to the usual conservation equation, i.e., $\dot{\rho}+3H\rho=0$; when $h(-{\sf T})=0$ that is, in {\sf GR} case and $g({\sf R})$ gravity. The next situation could occur when the two expressions in parentheses are equal, giving then the solution (\ref{specific}). Otherwise, there is a modified version of the conservation equation. Therefore, equation (\ref{relation-3}) completely depicts the behavior of the trace/matter density with respect to the scale factor. After determining ${\sf T}(a)$ via using equation (\ref{relation-3}), one can solve for the equations of motion (\ref{first}) and (\ref{second}) to obtain the scale factor and then discuss the cosmological consequences. More precisely, we can rewrite equation (\ref{relation-3}) as follows:
\begin{align}\label{relation-4}
\int_{{\sf T}_{0}}^{-{\sf T}} \frac{\kappa^{2}-\frac{3}{2}h'+h''{\sf T}}{{\sf T}\left(\kappa^{2} -h'\right)}d{\sf T}=-3\int_{a_{0}}^{a} d(\ln a),
\end{align}
where ${\sf T}_{0}$ and $a_{0}$ are the present values. Principally, the above relation can give the trace as a function of the scale factor, however, the left hand side integral may not easily be solved in general. In other words, even if the integration is carried out, finding the density as an explicit function of scale factor may not be possible, at least analytically. Such a situation can occur when the {\sf EoS} is not taken simply as that of a perfect fluid. However, from (\ref{relation-4}) for $h(-{\sf T})=\chi^{2} (-{\sf T})^{\alpha}$ and a dust fluid we obtain
\begin{align}\label{sol-noncon}
\left[\frac{\left(\rho -\alpha  \beta  \rho ^{\alpha }\right)^{2 \alpha -1}}{\rho }\right]^{\frac{1}{2 (\alpha -1)}}=Ca^{-3},
\end{align}
where $C$ is a constant of integration. As is seen, solution (\ref{sol-noncon}) may not be further simplified for arbitrary values of $\alpha$. For applications in the dynamical system approach which we shall utilize in later sections, we rewrite equation (\ref{relation-3}) in terms of the dimensionless variables, as follows:
\begin{align}\label{relation-5}
\frac{\dot{\rho}}{3H\rho}=-\frac{\Omega^{\sf dust}+x_{5}}{\Omega^{\sf dust}+\left(\frac{3}{2}+n\right)x_{5}}.
\end{align}
Equation (\ref{relation-5}) reduces to the standard form for $n=-1/2$, which 
corresponds to the conserved model~(\ref{specific}). It is noteworthy to briefly point out an 
important thermodynamical trait of conservation equation (\ref{relation-3}). As the author of~\cite{Harko14} has beautifully detailed this issue, one can write down the conservation equation in $f({\sf R},{\sf T})$ gravity in order to explain the matter creation process (we do not enter the details of calculations). That is, the following equations can be obtained:
\begin{align}
&\dot{\rho}+3(\rho+p)H=(\rho+p)\Gamma,\nonumber\\
&\Gamma=-\frac{\mathcal{F}({\sf R},{\sf T})}{\kappa^{2}+\mathcal{F}({\sf R},{\sf T})}\left[\frac{d\mathcal{F}({\sf R},{\sf T})}{dt}+\frac{1}{2}\frac{\dot{\rho}-\dot{p}}{\rho+p}\right]\nonumber,
\end{align}
where $\Gamma$ is the particle creation rate. This equation denotes no particle creation when $\mathcal{F}=0$, e.g., in the case of $f({\sf R},{\sf T})=g({\sf R})+\Lambda$ gravity, where $\Lambda$ is a cosmological constant. We can interpret this modified conservation equation as indicating an irreversible matter creation process caused by an irreversible energy flow from the gravitational field to the matter constituents. Briefly, in the context of $f({\sf R},{\sf T})$ gravity, it is possible that the spacetime transmutes to matter. Using the concepts of quantum cosmology, the particles which are created in this way may be in the form of some scalar particles (bosons) which specify the {\sf DM} content of the Universe. One can also rewrite the above conservation equation as follows:
\begin{align}
&\dot{\rho}+3(\rho+p+p_{\textrm{c}})H=0,\nonumber\\
&p_{\textrm{c}}=-\frac{\rho+p}{3}\frac{\Gamma}{H}\nonumber.
\end{align}
The above form of conservation equation suggests a new term that is called the creation pressure, $p_{\textrm{c}}$~\cite{Prigogine}. From thermodynamic point of view, this means that the coupling of geometry and matter generates (or can be translated as) a new pressure effect (which may even be considered as an effective bulk viscosity~\cite{Harko14}) that participates in a (gravitational) particle creation procedure. Therefore, unlike some other models (such as {\sf GR} and $g({\sf R})$ gravity) within which the problem of particle creation must be considered based on an admissible physical processes, in the framework of models which allow for coupling between matter and geometry (such as $f({\sf R},{\sf T})$ gravity) the irreversible thermodynamic processes are completely determined by the gravitational action.
In the next section, we shall use equation (\ref{relation-5}) in order to construct the dynamical system representation of equations (\ref{eom1}) and (\ref{eom2}) and study its cosmological consequences.
\section{Late-time behavior in $f({\sf R},{\sf T})=g({\sf R})+h({\sf T})$ gravity}\label{gen}
In this section we study cosmological consequences of models of the type $f({\sf R},{\sf T})=g({\sf R})+h({\sf T})$
allowing for the {\sf EMT} conservation to be violated, that is we suppose $\nabla_{\alpha}{\sf T}^{\alpha\beta}\neq0$. Equations (\ref{eom1}) and (\ref{eom2})
in terms of dimensionless variables (\ref{varx1})-(\ref{s}) are obtained:
\begin{align}
&\Omega^{\sf dust}+x_{1}+x_{2}+x_{3}+x_{4}+x_{5}=1,\label{gen1}\\
&\frac{\ddot{g}'}{Hg'}=1+2x_{1}+3x_{2}+x_{3}+3x_{4},\label{gen2}
\end{align}
respectively. Therefore, relations (\ref{relation-5})-(\ref{gen2}) would help us to obtain an autonomous system of equations
of motion for dimensionless variables $x_{1}-x_{5}$. Evolutionary equations are then obtained as follows:
\begin{align}
&\frac{d x_{1}}{d N}= -1+x_{1} (x_{1}-x_{3})-3x_{2}-x_{3}-3x_{4} \label{gen3},\\
&\frac{d x_{2}}{d N}= \frac{x_{1} x_{3}}{m} +x_{2}\left(4+x_{1} -2x_{3}\right)\label{gen4},\\
&\frac{d x_{3}}{d N}=- \frac{x_{1} x_{3}}{m} +2x_{3} \left(2 -x_{3}\right)\label{gen5},\\
&\frac{d x_{4}}{d N}=x_{4} (x_{1}-2x_{3}+4)-\frac{3 x_{5} (1-x_{1}-x_{2}-x_{3}-x_{4})}
{2 \left[1-x_{1}-x_{2}-x_{3}-x_{4}+\left(n+\frac{1}{2}\right) x_{5}\right]}\label{gen6},\\
&\frac{d x_{5}}{d N}=x_{5} \left(-\frac{3 (n+1) (1-x_{1}-x_{2}-x_{3}-x_{4})}{1-x_{1}-x_{2}-x_{3}-x_{4}+\left(n+\frac{1}{2}\right) x_{5}}+x_{1}-2x_{3}+4\right).\label{gen6}
\end{align}
As can be seen, both $n$ and $m$ parameters appear within the above system of equations, that is,
fixed point solutions and their properties depending on the parameters ($n$ and $m$) which exhibit the nature of the underlying model. In Table~\ref{tot} we have summarized equilibrium points and their cosmological features for the above system.
\begin{center}
\begin{table}[h!]
\centering
\caption{The fixed points solutions of $f({\sf R},{\sf T})=g({\sf R})+ h(-{\sf T})$ gravity.}
\begin{tabular}{l @{\hskip 0.1in} l@{\hskip 0.1in} l @{\hskip 0.1in}l@{\hskip 0.1in}l@{\hskip 0.1in}l}\hline\hline\\[-2 ex]
Point     &Coordinates $(x_{1},x_{2},x_{3},x_{4},x_{5})$&$w^{\textrm{(eff)}}$&$\Omega^{\sf dust}$&$\Omega^{\textrm{({\sf DE})}}$&\\[0.4 ex]
\hline\\[-2 ex]
${\sf P}_{1}$&$\left(-4,5 - x_{4},0,x_{4},x_{5}\right)$&$\frac{1}{3}$&$-x_{5}$&$1+x_{5}$\\[0.75 ex]
${\sf P}_{2}$&$\left(0,-1 - x_{4},2,x_{4},x_{5}\right)$&$-1$&$-x_{5}$&$1+x_{5}$\\[0.75 ex]
${\sf P}_{3}$&$\left(-4,\frac{4 (n+1) (5 n+9)}{(2 n+3)^2},0,\frac{4 n+9}{(2 n+3)^2},x_{5}\right)$&$\frac{1}{3}$&$-x_{5}$&$1+x_{5}$\\[0.75 ex]
${\sf P}_{4}$&$\left(0,-\frac{4 (m+1)^2 n^2+(m (4 m+21)+14) n+3 (5 m+4)}{(m+1)^2 (2 n+3)^2},2,\right.$\\
&$\left.\frac{-m (8 m n+9 m+3 n+3)+2 n+3}{(m+1)^2 (2 n+3)^2},x_{5}\right)$&$-1$&$-x_{5}$&$1+x_{5}$\\[0.75 ex]
${\sf P}_{5}$&$\left(-1,0,0,0,0\right)$&$\frac{1}{3}$&$2$&$-1$\\[0.75 ex]
${\sf P}_{6}$&$\left(\frac{6 m (n+1)}{(m+1) (2 n+3)},-\frac{m (4 n+6)+n+3}{(m+1)^2 (2 n+3)},\frac{m (4 n+6)+n+3}{(m+1) (2 n+3)},\right.$\\
&$\left.\frac{-m (8 m n+9 m+3 n+3)+2 n+3}{(m+1)^2 (2 n+3)^2},\right.$\\
&$\left.-\frac{2 (n+1) (m (8 m n+9 m+3 n+3)-2 n-3)}{(m+1)^2 (2 n+3)^2}\right)$&$-\frac{m (2 n+3)+1}{(m+1) (2 n+3)}$&$0$&$1$\\[0.75 ex]
${\sf P}_{7}$&$\left(-\frac{2 (n+3)}{2 n+3},0,0,\frac{4 n+9}{(2 n+3)^2},\frac{2 (n+1) (4 n+9)}{(2 n+3)^2}\right)$&$\frac{1}{3}$&$0$&$1$\\[0.75 ex]
${\sf P}_{8}$&$\left(\frac{3 m}{m+1},-\frac{4 m+1}{2 (m+1)^2},2-\frac{3}{2 (m+1)},0,0\right)$&$-1+\frac{1}{m+1}$&$\frac{2-m (8 m+3)}{2 (m+1)^2}$&$\frac{m (10 m+7)}{2 (m+1)^2}$\\[0.75 ex]
\hline\hline
\end{tabular}
\label{tot}
\end{table}
\end{center}
Before discussing the solutions and their stability properties, there are some issues that should be pointed out.  Note that, since the radiation fluid is absent in the present study, we exclude fixed points with $w^{({\sf eff})}=1/3$ from our considerations. As Table~\ref{tot} indicates, there are some fixed points for which the coordinates and cosmological features are not completely determined. More precisely, fixed point ${\sf P}_{2}$ is the solution of a set of equations (\ref{gen3})-(\ref{gen6}) for arbitrary values of $x_{4}$ and $x_{5}$  and ${\sf P}_{4}$ is a solution for arbitrary values of $x_{5}$. Naturally, the question may arise of whether it is possible to determine exactly the physical properties of the fixed points. The answer is positive! There are some mathematical and physical criteria that must be met. Since at the critical points we have $dx_{i}/dN=0,~i=1,\cdots,5$, in general, this condition must hold for every function, namely $\mathcal{W}(x_{1},x_{2},x_{3},x_{4},x_{5})$. For instance, $r=r(x_{2}, x_{3})$ and $s=s(x_{4}, x_{5})$ must be stationary at the fixed point. This means that
\begin{align}\label{gen7}
\frac{dr}{dN}=\frac{\partial r(x_{2}, x_{3})}{\partial x_{2}}\frac{dx_{2}}{dN}
+\frac{\partial r(x_{2}, x_{3})}{\partial x_{3}}\frac{dx_{3}}{dN}=0,
\end{align}
and
\begin{align}\label{gen8}
\frac{ds}{dN}=\frac{\partial s(x_{4}, x_{6})}{\partial x_{4}}\frac{dx_{4}}{dN}
+\frac{\partial s(x_{4}, x_{6})}{\partial x_{6}}\frac{dx_{6}}{dN}=0,
\end{align}
whereby using definitions (\ref{varx2})--(\ref{varx5}) and (\ref{m})--(\ref{s}), we have
\begin{align}\label{gen9}
0=\frac{d r}{d N}=-r\left(\frac{1+r+m(r)}{m(r)}\right) x_{1}\equiv-r {\mathcal M} (r) x_{1},
\end{align}
and
\begin{align}\label{gen10}
0=\frac{d s}{d N}=3s\Big{(}s-n(s)-1\Big{)},
\end{align}
where we have defined
\begin{align}\label{gen11}
{\mathcal M} (r)\equiv \frac{1+r+m(r)}{m(r)}.
\end{align}
As a result, solutions must accept the conditions $m=-r-1$ (provided that $m(r)\neq0$), $r=0$ or $x_{1}=0$ and $n=s-1$ or $s=0$. The condition $r=x_{3}/x_{2}=0$ leads to $x_{3}=0$ and condition $s=x_{5}/2x_{4}=0$ gives $x_{5}=0$. This fact restricts the value of $x_{5}$ to null, and with this choice the location of point ${\sf P}_{4}$ in the phase space will be completely determined. Moreover, ${\sf P}_{2}$ and ${\sf P}_{4}$ must have $x_{5}=0$ in order to completely represent a dominant de Sitter phase. Therefore, one of the novel features developed by violation of {\sf EMT} conservation is the appearance of a new de Sitter solution (${\sf P}_{4}$) as compared to the case of pure $f({\sf R})$ gravity, for which the properties depend upon the matter sector of the Lagrangian.\footnote{$f({\sf R})$ gravity theories accept a de Sitter solution in the late times with coordinates $(0,-1,2)$ as authors of \cite{amend} have shown.} Another interesting result is that a {\sf DE} fixed point (${\sf P}_{6}$) solution is achieved within this framework. This point indicates a dominant {\sf DE} solution with an effective equation of state which depends on both the geometrical and the matter part of the Lagrangian via the parameters $m$ and $n$. Let us now proceed with exploring the stability properties of the fixed points ${\sf P}_{2}$, ${\sf P}_{4}$, ${\sf P}_{6}$ and ${\sf P}_{8}$.
\begin{itemize}
\item The equilibrium points ${\sf P}_{2}$ and ${\sf P}_{4}$\vspace{.3cm}\\
These two points have the same eigenvalues given as
\begin{align}\label{gen12}
-\frac{5}{2},~-3 ,~-\frac{6 n}{2 n+1},~-\frac{1}{2} \left(3+\sqrt{25-\frac{16}{m}}\right),~-\frac{1}{2} \left(3-\sqrt{25-\frac{16}{m}}\right).
\end{align}
Therefore, these points are representative of stable equilibrium points if $m$ and $n$ parameters satisfy the following intervals\footnote{Indeed, the last two eigenvalues given in (\ref{gen12}) will become negative for $\frac{16}{25}\leqslant m<1$ signaling that these two points are stable nodes. While, for $0<m<\frac{16}{25}$, the last two eigenvalues become complex with pure negative real part for which the corresponding point is called stable focus in the literature. However, these type of fixed points are asymptotically stable.}
\begin{align}\label{gen13}
0< m<1,~~~\mbox{with}~~~n<-\frac{1}{2}~~\mbox{or}~~ n>0.
\end{align}
The stability region for these solutions has been presented in gray color in the right panel of Figure~\ref{fig1}. I is seen that, for every curve lying within the gray region, ${\sf P}_{2}$ or ${\sf P}_{4}$ will represent a late time solution.
\item The equilibrium Point ${\sf P}_{6}$\vspace{.3cm}\\
The eigenvalues for this equilibrium point are obtained as
\begin{align}\label{gen14}
\frac{6 n}{2 n+3},~i(m,n),~j(m,n),~\frac{6 (n+1) \left(m'+1\right)}{2 n+3},~-\frac{6 (n+1) \left(n'-1\right)}{2 n+3},
\end{align}
where
\begin{align}\label{gen14}
&i(m,n)=-\frac{3 \sqrt{m} (m+n+2)+k(m,n)}{2 \sqrt{m} (m+1) (2 n+3)},\nonumber\\
&j(m,n)=\frac{-3 \sqrt{m} (m+n+2)+k(m,n)}{2 \sqrt{m} (m+1) (2 n+3)},\\
&k(m,n)=\Bigg{[}m^3 (16 n+21)^2+2 m^2 (5 n+6) (16 n+33)\nonumber\\
&-m (n (31 n+60)+36)-8 (n+3) (2 n+3)\Bigg{]}^{1/2},\nonumber
\end{align}
where $m'\equiv dm(r)/dr$ and $n'\equiv dn(s)/ds$ must be calculated at the desired fixed point. There are two class of solutions in which the fixed point ${\sf P}_{6}$ is a stable point. The first class corresponds to those for which ${\sf P}_{6}$  is a stable node and those that ${\sf P}_{6}$ is a spiral source. We have plotted the stability region for this point in the left panel of Figure~\ref{fig1}.
\begin{figure}[h!]
\centering
\centerline{\epsfig{figure=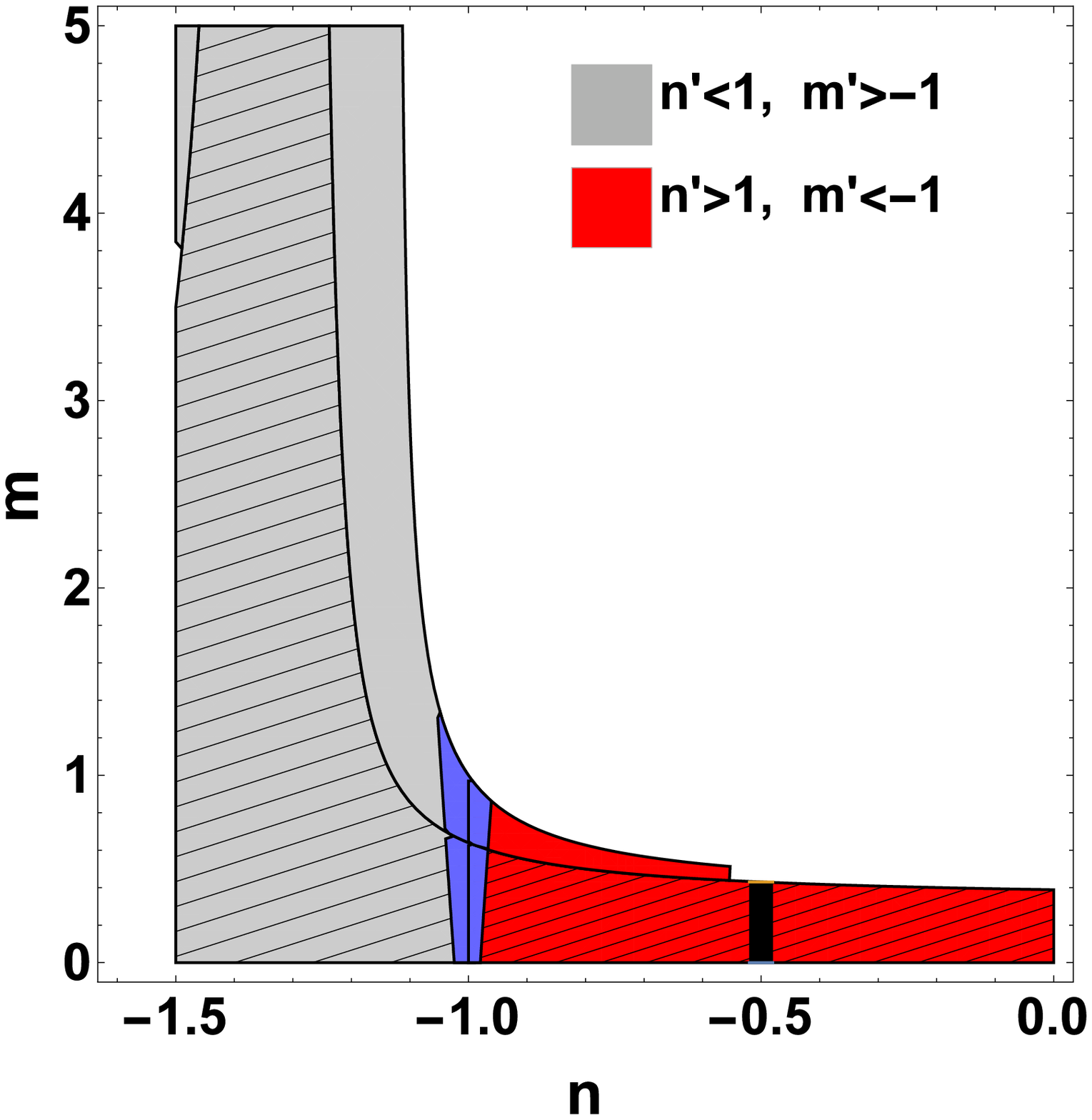,width=6cm}\hspace{2mm}\epsfig{figure=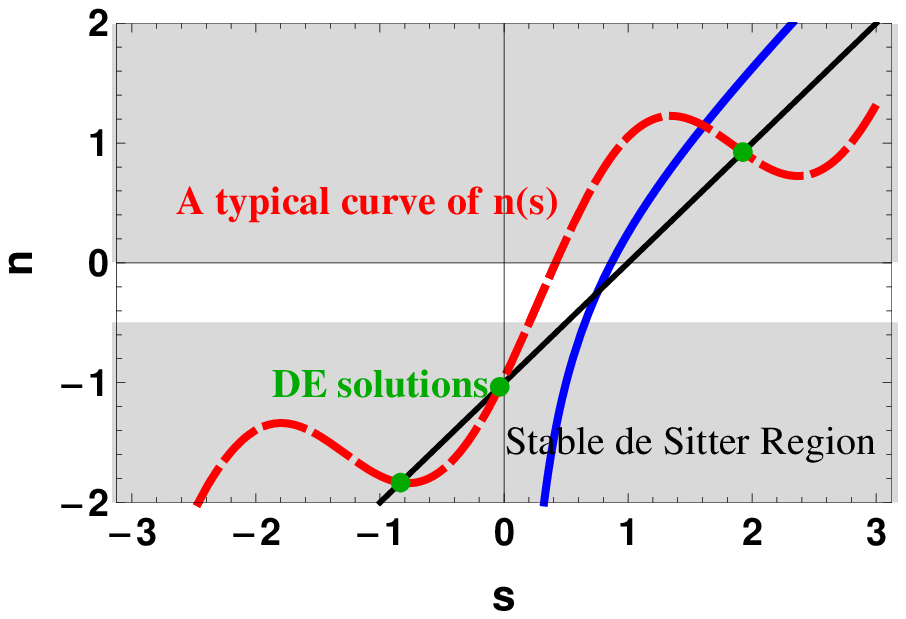,width=7cm}}
\caption{Left panel: the stability region of fixed point ${\sf P}_{6}$. Hatched areas show spiral stability. Black region which occurred in $n=-1/2$ indicates situation of the conserved model $f({\sf R},{\sf T})=g({\sf R})+\sqrt{-{\sf T}}$. Blue colored regions show solutions which are compatible with Planck 2015 data for dark energy equation of state. Right Panel: The gray areas show the regions in which de Sitter solutions are stable. The black line represents the line $n=s-1$ and the green points indicate dark energy solutions which correspond to ${\sf P}_{6}$ fixed points for hypothetical roots $s_{u}$. These roots are intersection of $n(s)$ of model (the red curve) with the line $n=s-1$. Blue line shows curve $n(s)$ for model with $h(-{\sf T})=c_2 {\rm exp}[{-c_1 (-{\sf T})-\beta (-{\sf T})^4/12}]$, which is discussed in Section~\ref{gen}.}
\label{fig1}
\end{figure}

Point ${\sf P}_{6}$ is the only critical point which lies both on the $m=-r-1$ and $n=s-1$ curves, that is for this fixed point we have $x_{3}/x_{2}=r=(-1-m)$ and $x_{5}/2x_{4}=s=n+1$. This means that, ${\sf P}_{6}$ is a {\sf DE} solution only for the roots of equations $n(s_{u})=s_{u}-1,~u=1,...,N$ and $m(r_{v})=r_{v}-1,~v=1,...,\mathcal{N}$. In other words, if for a model specified in terms of the functions $m(r)$ and $n(s)$, there exit a class of $s_{u}$ and $r_{v}$ roots, we then can conclude that the model would accept ${\sf P}_{6}$ as a {\sf DE} solution at late times. Besides, conditions on $n'$ or $m'$ as shown in the left panel of Figure~\ref{fig1} must be satisfied for these roots. For example, for a model admitting an $s_{u}$ and $r_{v}$ solution (in blue colored region in Figure~\ref{fig1}) for which either $n'(s_{u})<1, m'(r_{v})>-1$ or $n'(s_{u})>1, m'(r_{v})<-1$ holds, a {\sf DE} solution consistent with Plank 2015 data exists.

\item The equilibrium Point ${\sf P}_{8}$\vspace{0.3cm}\\
By considering the matter density and {\sf EoS} parameters for this fixed point, we find out that it gives a matter dominated solution in the limit $m\to0$. The eigenvalues for this point are obtained as
\begin{align}\label{gen15}
&-\frac{3+\sqrt{ 32 m (8 m+5)-31-\frac{16}{m}}}{4 (m+1)},~\frac{-3+\sqrt{32 m (8 m+5)-31-\frac{16}{m}}}{4 (m+1)},\nonumber\\
&-3 n,~3 \left(m'+1\right),~3-3 (n+1) n'.
\end{align}
This fixed point is unstable under the following conditions
\begin{align}\label{gen16}
&n<0,~~n'=0,~~n'<0~\mbox{with}~ n>\frac{1-n'}{n'},~~0<n'<1~ \mbox{with}~ n<\frac{1-n'}{n'},\nonumber\\
&m<\frac{1}{16} \left(-3-\sqrt{73}\right),~~-\frac{1}{4}<m<0,~~m>\frac{1}{16} \left(\sqrt{73}-3\right),~~m'>-1.
\end{align}
We note that, in order to avoid a short matter dominated era or even the absence of it, we must chose the limit $m\to0^{+}$. For $m\to0^{-}$, the second eigenvalue becomes a large positive number and hence, in this case ${\sf P}_{8}$ indicates a very transient matter dominated era. Numerical manipulations show that, in order to have a proper matter era, that is for $m\to0^{+}$, the condition $m'>-1$ must be satisfied, otherwise, matter dominated solutions would be disappeared. Actually, numerical studies reveal that, for models with a ${\sf P}_{6}$ solution lying within the red area in the left panel of Figure~\ref{fig1}, there is no matter dominated solution. There is only a direct transition from either of the critical points with $w^{({\sf eff})}=1/3$ to ${\sf P}_{2}, {\sf P}_{4}$ or ${\sf P}_{6}$.
\end{itemize}

\noindent
Mathematically accepted cosmological solutions are those that include a true connection between the matter dominated fixed point and late time solutions which correspond to de Sitter points ${\sf P}_{2}$ or ${\sf P}_{4}$ or the {\sf DE} fixed point ${\sf P}_{6}$. Thus, the possible cosmological solutions are classified as follows:

\begin{itemize}
\item Solutions which correspond to trajectories from ${\sf P}_{8}$ with $m'(r\to-1^{-})>-1$ to ${\sf P}_{2}$ or ${\sf P}_{4}$ with $n>0$ or $n<-1/2$.\\
For a certain model which is specified by its $m(r)$ and $n(s)$ functions, physically justified transitions in
phase space occur for those conditions which make ${\sf P}_{8}$ an unstable fixed point and ${\sf P}_{2}$ or
${\sf P}_{4}$ a stable one, simultaneously. From the right panel of Figure~\ref{fig1}, there are two stable regions for ${\sf P}_{2}$ and ${\sf P}_{4}$; The regions are defined by intervals $0<m(r_{v})<1,~n(s_{u})<-1/2$ and $0<m(r_{v})<1,~n(s_{u})>0$
for the roots $r_{v}$ and $s_{u}$. Since, the models must be designed in such a way that the condition $m'(r\to-1^{-})>-1$ holds, all de Sitter solutions can be connected to matter ones. In the region with $n(s)<0$, cosmological solutions with a de Sitter phase at late times can be classified as two distinct categories; {\bf a)} initial values can be set such that ${\sf P}_{6}$ solution would not be satisfied. In this case, there is only a direct transition from matter to de Sitter solution. {\bf b)} However, initial values can also be chosen so that an unstable ${\sf P}_{6}$ solution be realized. In such cases an intermediate transient ${\sf P}_{6}$ phase for $-0.5<n<0$\footnote{See the white narrow band in the right panel of Figure~\ref{fig1}.} would appear. To illustrate these two types of solutions we have drawn in Figure~\ref{fig3}, the effective {\sf EoS} parameter for a class of models with $g({\sf R})={\sf R}\log[\alpha {\sf R}]$ which corresponds to $m(r)=r+1/r$, and $h(-{\sf T})=c_2 {\rm exp}{[-c_1 (-{\sf T})-\beta (-{\sf T})^4/12]}$ giving $n(s)=s-\beta/s$, for $\beta=0.75$ and two different sets of initial values for the variables $x_{1}-x_{5}$\footnote{The effective equation of state in figure~\ref{fig3} is independent of the constants $\alpha$, $c_{1}$ and $c_{2}$.}. In this example, we always have $m'(r)>-1$ and $n'(s)>1$ leading to an unstable ${\sf P}_{6}$ fixed point which is recognized by the root $s_{v}=\beta=0.75$ for which $n(0.75)=-0.25$ and correspondingly gives $w^{\rm{({\sf DE})}}_{0}=-0.4$. Thus, the $n(s)$ curve evolves into $n<-1/2$ or $n>0$ regions in which the de Sitter fixed point is stable (see blue curve in the right panel of Figure~\ref{fig1}).

For models with $n>0$ there is a proper cosmological solution corresponding to a direct transition from matter to de Sitter phases.
\begin{figure}[h!]
\centering
\centerline{\epsfig{figure=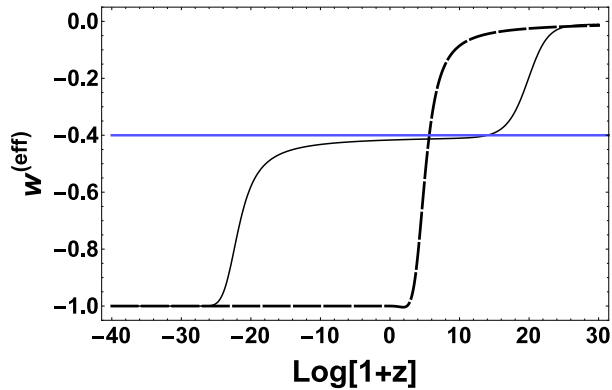,width=8cm}\hspace{2mm}}
\caption{The effective equation of state for two different cosmological solutions with de Sitter phase for models with $g({\sf R})={\sf R}\log[\alpha {\sf R}]$ and $h(-{\sf T})=c_2 {\rm exp}{[-c_1 (-{\sf T})-\beta (-{\sf T})^4/12]}$. Dashed curve shows a direct transition from matter era to de Sitter phase. Full curve indicates an intermediate ${\sf P}_{6}$ solution.}
\label{fig3}
\end{figure}
\item Solutions which correspond to trajectories from ${\sf P}_{8}$ with $m'(r\to-1^{-})>-1$ to ${\sf P}_{6}$.\\
Figure~\ref{fig1} shows that a necessary stability condition for ${\sf P}_{6}$ is $-3/2<n<0$. However, as it is already emphasized, there is no proper matter dominated solution in the region where $-1<n<0$. Models with $n<-1$ has
been already accepted as de Sitter solution. This is because the curve $n(s)$ for all values approaching to a ${\sf P}_{6}$ root in the region $-3/2<n(s_{v})<-1$, lies within the stable de Sitter range. Briefly speaking, there is no proper cosmological solution with a {\sf DE} phase at late times which corresponds to a ${\sf P}_{6}$ solution.
\end{itemize}

There are some classes of minimal models for which the above theme slightly differs; models that each of gravitational or matter sectors or both of them within the Lagrangian are in the form of power law. In these cases, $m'$ or $n'$ or both vanish, which means that the pairs ($m,r$) or ($n,s$) or both of them are constant. Hence, since $x_{3}=rx_{2}$ and $x_{5}=2sx_{4}$, the number of independent variables would reduce from five to four or three. This fact leads to some changes in the fixed point solutions and their stability of the dimensionally reduced dynamical system. To illustrate this issue, we shall investigate the class of models for which both parts of their Lagrangian are power law. This is the subject of next section.
\section{Pure minimal case $f({\sf R},{\sf T})={\sf R}^{\alpha}+(-{\sf T})^{\beta}$}\label{pure}
In this section, we briefly study the late time solutions of pure minimally coupled models of the type $f({\sf R},{\sf T})={\sf R}^{\alpha}+(-{\sf T})^{\beta}$. For these models we obtain $r=x_{3}/x_{2}=-\alpha$  and $s=x_{5}/2x_{4}=\beta$. That is, for these cases, there are only three independent variables $x_{1}$, $x_{2}$ and $x_{4}$. Hence, this dimensional reduction demands slightly different method. Here, because $m'=0$ and $n'=0$, the stability properties of fixed points are described by the $m$ and $n$ parameters only. For these types of models there are six fixed point solutions of which three are of physical interest, as shown in Table~\ref{tab2}.
\begin{center}
\begin{table}[h!]
\centering
\caption{The fixed points solutions of $f({\sf R},{\sf T})={\sf R}^{\alpha}+ (-{\sf T})^{\beta}$ gravity.}
\begin{tabular}{l @{\hskip 0.1in} l@{\hskip 0.1in} l @{\hskip 0.1in}l@{\hskip 0.1in}l@{\hskip 0.1in}l}\hline\hline\\[-2 ex]
Point     &Coordinates $(x_{1},x_{2},x_{4})$&$w^{({\sf eff})}$&$\Omega^{\sf dust}$&$\Omega^{\textrm{({\sf DE})}}$&\\[0.4 ex]
\hline\\[-2 ex]
${\sf P}^{\rm{({\sf M})}}$&$\left(\frac{3 m}{m+1},-\frac{4 m+1}{2 (m+1)^2},0\right)$&$\frac{m (10 m+7)}{2 (m+1)^2}$&$\frac{2-m (8 m+3)}{2 (m+1)^2}$&$\frac{1}{m+1}-1$\\[0.75 ex]
${\sf P}^{\rm{({\sf S})}}$&$\left(0,\frac{2}{m+1},\frac{2}{m+1}-1\right)$&$-1$&$\frac{2 (m-1) (n+1)}{m+1}$&$\frac{-2 m n-m+2 n+3}{m+1}$\\[0.75 ex]
${\sf P}^{\rm{({\sf DE})}}$&$\left(\frac{6 m (n+1)}{(m+1) (2 n+3)},-\frac{m (4 n+6)+n+3}{(m+1)^2 (2 n+3)},\right.$\\
&$\left.\frac{-m (8 m n+9 m+3 n+3)+2 n+3}{(m+1)^2 (2 n+3)^2}\right)$&$-\frac{m (2 n+3)+1}{(m+1) (2 n+3)}$&$0$&$1$\\[0.75 ex]
\hline\hline
\end{tabular}
\label{tab2}
\end{table}
\end{center}
 It is interesting to note that a de Sitter and a {\sf DE} solution have appeared again. The first fixed point can denote the matter dominated era only for $m=(\alpha-1)\to0^{+}$ which leads to $\alpha\to 1^{+}$. Note that the eigenvalues of this point are given by the first three parts of expression (\ref{gen15}), that is, the matter dominated point is well behaved only for $m\to 0^{+}$ and is unstable only when $n<0$ as the other two eigenvalues are negative for $m\to 0^{+}$.
The stability conditions for de Sitter point ${\sf P}^{{\sf (S)}}$ are the same as (\ref{gen13}) which means that only for $n<-1/2$, the matter dominated point ${\sf P}^{{(\sf M)}}$ can be connected to de Sitter point. Since the matter density of a de Sitter fixed point depends on the model parameters, this fact implies that to have a thoroughly de Sitter dominant phase (i.e., $\Omega^{\sf dust}=0$ and $\Omega^{\rm{({\sf DE})}}$=1) we must have $n\to-1$ for arbitrary non-zero $m$. Therefore, the model with $m\to0^{+}$ and
$n=-1$ includes a true connection from a long enough matter dominated era to a stable de Sitter era at late times. The value $n\to-1$ corresponds to $\beta=0$, from which in this case we have a cosmological constant instead of the function $h(-{\sf T})$.

\begin{figure}[h!]
\centering
\centerline{\epsfig{figure=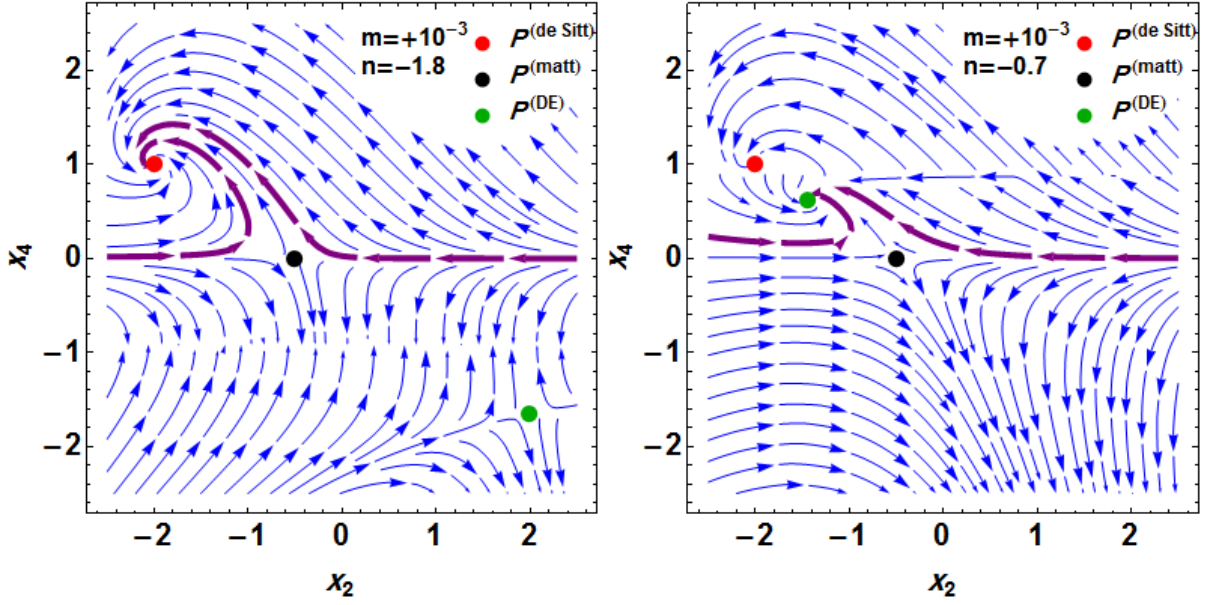,width=16cm}}
\caption{The situation of the two fixed points, i.e., {\sf DE} and de Sitter solutions for the model $f({\sf R},{\sf T})={\sf R}^{\alpha}+(-{\sf T})^{\beta}$ and for two different values of $n$ parameter. Left panel: The green point represents an unstable {\sf DE} solution. Right panel: The green point represents a stable {\sf DE} solution. The trajectories passing through the matter dominated fixed point will be terminated at the {\sf DE} solution before reaching the de Sitter point.}
\label{fig2}
\end{figure}

The eigenvalues for ${\sf P}^{^{\rm{({\sf DE})}}}$ are given by the first three parts given in expression (\ref{gen14}) which show that this point is stable within the same region as shown in Figure~\ref{fig1}, regardless of the the conditions on $m'$ and $n'$. Therefore, all models with $m>0$ and
$-3/2<n<0$ could provide a true connection from matter dominated to {\sf DE} fixed points. However, the blue colored zone in  Figure~\ref{fig1} restricts models to those that are consistent with the observational data. This means that, only models with $n\to-1$ are acceptable. To show the behavior of such a model with two late time solutions we have plotted in Figure~\ref{fig2}, the phase space diagrams for two different values $n=-0.7,-1.8$ with $m=+0.001$ in $(x_{2},x_{4})$ plane. The red point denotes the de Sitter fixed point, black one shows the matter dominated point and the green point indicates the {\sf DE} one. The purple trajectories show physically justified connections. As can be seen, the {\sf DE} point is unstable for $n<-3/2$ which makes ${\sf P}^{{\sf (S)}}$ as the only late time solution. For $-3/2<n<-1/2$ both ${\sf P}^{{\sf (S)}}$ and ${\sf P}^{^{\rm{({\sf DE})}}}$ are stable, nevertheless, for $-1<n<-1/2$ the trajectories will be trapped by ${\sf P}^{^{\rm{({\sf DE})}}}$ (see Figure~\ref{fig2}) and for $n<-1$ will be attracted by ${\sf P}^{^{\sf (S)}}$. For $n=-1$ the two points coincide. A particular case has occurred; the coordinates of the {\sf DE} fixed point and its eigenvalues (see the results given in (\ref{gen14})) are singular for $n=-3/2$. In~\cite{fRT12}, we have discussed in detail the cosmological behavior of this case by algebraic treatments and have shown that in this case the late time solution corresponds to a de Sitter era. As a final remark, we point out that, from 2015 Planck data~\cite{Planck} which has determined the present value of the {\sf EoS} of {\sf DE} as $-1.051<w_{0}^{\rm{({\sf DE})}}<-0.961$, we may conclude that power law models can be admissible only for $m\to0^{+}$ and $-1.025<n<-0.980$ which corresponds to $-0.024<\alpha<0.02$. Therefore, these types of modified gravity theories, by including two different late time solutions, deserve more investigations as alternatives of the $\Lambda$CDM model.
\section{Concluding remarks}\label{con}
In this work, in the context of a particular class of $f({\sf R},{\sf T})$ gravity theories, we have discussed the issue of accelerated expansion of the Universe at the late time cosmological regimes under assumption that the {\sf EMT} conservation is violated. Though the assumption on relaxing the {\sf EMT} conservation may not necessarily provide a late time accelerated expansion scenario, we tried to seek for possible relations between such an assumption and the cosmic late time acceleration within the framework of $f({{\sf R},{\sf T}}$) gravity. Specifically, our attention was concentrated on those models for which the Lagrangian can be chosen with minimal coupling between the geometrical and matter sectors, which we call minimally coupled models and which can be written as $f({\sf R},{\sf T})=g({\sf R})+h({\sf T})$. We utilized the dynamical system approach in order to reformulate the field equations. We used a way of parameterizations of models via introducing the parameters $m={\sf R} g''({\sf R})/g'({\sf R})$ and $r=-{\sf R} g'({\sf R})/g({\sf R})$  for the geometrical part of the Lagrangian and $n=-{\sf T} h''(-{\sf T})/h(-{\sf T})$ and $s=-{\sf T}h'(-{\sf T})/h(-{\sf T})$ for the matter part. Thus, an $f({\sf R},{\sf T})$ model can be determined by $m(r)$ and $n(s)$ functions. These parameters appear within the coordinates of equilibrium points of the system and also within their corresponding eigenvalues. Therefore, all cosmological properties of the system, as derived through exploring the eigenvalues, will depend upon these parameters. In fact, these parameters play a fundamental role for determining the cosmological viability of $f({\sf R},{\sf T})$ models under relaxing the condition on {\sf EMT} conservation. The application of this approach to explain the late time cosmological acceleration has been introduced in~\cite{amend} in the framework of $f(\sf{R})$ gravity models. It has also been used in $f({\sf R},{\sf T})={\sf R}+\alpha{\sqrt{\sf T}}$ as the only case which respects the conservation of {\sf EMT}~\cite{fRT2} and within the framework of $f({\sf R},{\sf T})={\sf R}+\alpha\Lambda(\sf {T})$ gravity~\cite{fRT12}, to extract cosmological solutions. Therefore, the current study may be regarded as an attempt to complete the previous investigations on the cosmological late time behavior of the minimally coupled $f({\sf R},{\sf T})$ gravity models. The study of the dynamical system representation of the field equations shows that there are three fixed points that can represent the late time behavior of the model; two de Sitter points together with a critical point for which $\Omega^{\rm{({\sf DE})}}=1$ and $\Omega^{\sf{(m)}}=0$ and its effective {\sf EoS} depends on the mentioned parameters. We call the latter fixed point the {\sf DE} solution. Having investigated and analyzed the properties of fixed points, we found that three types of transition between matter to either of de Sitter or {\sf DE} fixed points can occur. Therefore, depending on the underlying model as well as the initial values, there are three classes of cosmological solutions, so that, for all cases, in order to have a prolonged matter dominated era the condition $m(r\lessapprox-1)=0^{+}$ must hold. One can construct models in such a way that a \lq\lq{}{secure}\rq\rq{} transition from the matter to a de Sitter solution occurs. These types of models have to satisfy the condition $n>0$. Some models include a {\sf DE} fixed point at late times if the conditions $-1/2<n<0$, $n'>1$ and $m'(r\lessapprox-1)<-1$ are fulfilled. However, in these models the effective {\sf EoS} does not match the observational data. Finally, in some models, before reaching the de Sitter phase there is a transient phase of the {\sf DE} solution, as well. All models of this class must respect the condition $n<-1/2$. Note that, in the last two cases, one could set the initial values in order to obtain a direct transition to a de Sitter phase. However, care must be taken; since for the de Sitter fixed points the matter and {\sf DE} densities depend on the fixed point coordinates, the constants of the model must be chosen such that they lead to a dominant de Sitter phase at late times.

In summary, we can conclude that violation of {\sf EMT} conservation in the minimal $f({\sf R},{\sf T})$ gravity models could provide conditions under which the underlying model includes a de Sitter expansion at late times, at least for some subset of space parameters of the model. This fact indicates that allowing for the violation of {\sf EMT} conservation, the situation of minimal $f({\sf R},{\sf T})$ gravity theories can be encouraging enough for cosmological models and merits further investigation, in comparison with the same models when {\sf EMT} conservation is taken for granted. As the authors of~\cite{fRT2,fRT4,fRT14} have already elucidated, respecting the {\sf EMT} conservation leads to a {\sf DE} description at late times but the effective {\sf EoS} for such models would be found as $w^{({\sf eff})}=-1/2$, which could not be observationally accepted.

 We also have considered models with power law sections within the Lagrangian as $f({\sf R},{\sf T})={\sf R}^{\alpha}+(-{\sf T})^{\beta}$. It is found that these types of models can describe the recently accelerated expansion of the Universe for $\alpha\to1^{+}$ and $-0.024<\beta<0.02$.

As the final remark, we would like to highlight that the problem of de Sitter solution in the late time and more importantly the transition from matter dominated era to de Sitter era has been investigated previously in some other work. However, most was based on a reconstruction method in order to obtain models which include the desired behaviors~\cite{Jamil12,Houndjo12,Singh14,Sharif141}. Nevertheless, the method presented in the herein work may provide a more comprehensive plan to investigate all late time solutions as well as their stability.


\begin{thebibliography}{99}
\bibitem{supno3}  Riess, A.G., {\it et al.} ``BVRI curves for $22$ type Ia supernovae", \textit{Astron. J.} \textbf{117} (1999), 707.
\bibitem{SDSS2}   Abazajian, K., {\it et al.} ``The third data release of the Sloan Digital Sky Survey ", \textit{Astron. J.} \textbf{129} (2005), 1755.
\bibitem{WMAP5} Hinshaw, G. F., {\it et al.} ``Nine-Year wilkinson microwave anisotropy probe (WMAP) observations: cosmological parameter results", \textit{Astrophys. J. Suppl.} \textbf{208} (2013), 19.
\bibitem{Planck} Ade, P. A. R., {\it et al.} ``Planck 2015 results. XIII. cosmological parameters", \textit{A}\&\textit{A} \textbf{594} (2016), A13.
\bibitem{LCDM}     Ostriker, J. P. \& Steinhardt, P. J., \lq\lq{}Cosmic concordance\rq\rq{}, \textit{astro-ph/9505066}.
\bibitem{Cos.pro1} Weinberg, S., ``The cosmological constant problem'', \textit{Rev. Mod. Phys.} \textbf{61} (1989), 1.
\bibitem{Cos.pro3} Padmanabhan, H. \& Padmanabhan, T., ``CosMIn: The solution to the cosmological constant problem'',
\textit{Int. J. Mod. Phys. D} \textbf{22} (2013), 1342001.
\bibitem{darkmatt} Strigari,  L. E., \lq\lq{}Galactic searches for dark matter,\rq\rq{} {\it Phys. Rep.} {\bf 531} (2013), 1.
\bibitem{DMsector} Copeland, E. J., Sami, M. and Tsujikawa, S.,\lq\lq{}Dynamics of dark energy,\rq\rq{} {\it Int. J. Mod. Phys. D} {\bf 15} (2006), 1753;\\ Cai, Y. -F., Saridakis E. N., Setare M. R. and Xia J. -Q., \lq\lq{}Quintom cosmology: Theoretical implications and observations,\rq\rq{} {\it Phys. Rept.} {\bf 493} (2010), 1.
\bibitem{gravitsect} Capozziello, S. and Laurentis, M. De\lq\lq{}Extended Theories of Gravity,\rq\rq{} {\it Phys. Rept.} {\bf 509} (2011), 167.
\bibitem{fR} De Felice, A. \& Tsujikawa, S. \lq\lq{}$f({\sf R})$ theories\rq\rq{}, \textit{Living Rev. Rel.} \textbf{13} (2010), 3.
\bibitem{grsect100} Nojiri, S. i. and Odintsov, S. D., \lq\lq{}Unified cosmic history in modified gravity: From image theory to Lorentz non-invariant models,\rq\rq{} Phys. Rept. {\bf 505} (2011), 59.
\bibitem{grsect101} Lobo, F. S. N., \lq\lq{}The dark side of gravity: Modified theories of gravity,\rq\rq{} {\it arXiv:0807.1640 [gr-qc].}
\bibitem{saridynsys} Leon, G. \& Saridakis, E. N.\lq\lq{}Phase-space analysis of Horava-Lifshitz cosmology,\rq\rq{} {\it JCAP} 0911 (2009), 006; \lq\lq{}Dynamics of the anisotropic Kantowsky-Sachs geometries in $R^n$ gravity,\rq\rq{} {\it Class. Quant. Grav.} {\bf 28} (2011), 065008 ; Xu, C., Saridakis, E. N. and Leond, G.\lq\lq{}Phase-space analysis of teleparallel dark energy,\rq\rq{} {\it JCAP} 07 (2012), 005;\lq\lq{}Dynamical analysis of generalized Galileon cosmology,\rq\rq{} {\it JCAP} 1303 (2013), 025;\rq\rq{}Dynamical behavior in mimetic F(R) gravity,\rq\rq{} {\it JCAP} 1504 (2015), 031; \lq\lq{}Cosmology in time asymmetric extensions of general relativity,\rq\rq{}{\it JCAP} 1511 (2015), 11 009;\\Leon, G., Saavedra, J. and Saridakis, E. N. \lq\lq{}Cosmological behavior in extended nonlinear massive gravity,\rq\rq{}{\it  Class. Quant. Grav.} {\bf 30} (2013), 135001; \\Fadragas, C. R., Leon, G.  and Saridakis, E. N.\lq\lq{}Dynamical analysis of anisotropic scalar-field cosmologies for a wide range of potentials,\rq\rq{} {\it Class. Quantum Grav.} {\bf 31} (2014), 075018;\\ Kofinas, G., Leon G. and Saridakis, E. N.,\lq\lq{}Dynamical behavior in $f({\sf T},{\sf T}_G)$ cosmology,\rq\rq{} {\it Class. Quantum Grav.} {\bf 31} (2014), 175011; \\Skugoreva, M., Saridakis, E. N. and Toporensky, A.,\lq\lq{}Dynamical features of scalar-torsion theories,\rq\rq{} {\it Phys. Rev. D} {\bf 91} (2015), 044023;\\Carloni, S., Lobo, F. S. N., Otalora, G. and Saridakis, E. N.,\lq\lq{}Dynamical system analysis for nonminimal torsion-matter coupled gravity,\rq\rq{} {\it Phys. Rev. D} 93 (2016), 024034.
\bibitem{CLFEPASK} Clifton, T., Ferreira, P. G., Padilla, A., and Skordis, C., \lq\lq{}Modified gravity and cosmology,\rq\rq{} {\it Phys. Rep.} {\bf 513} (2012), 1.
\bibitem{fRT1} Harko, T., Lobo, F. S. N., Nojiri, S. \& Odintsov, S. D., \lq\lq{}$f({\sf R},{\sf T})$ gravity\rq\rq{}, \textit{Phys. Rev. D} \textbf{84} (2011), 024020.
\bibitem{Harko14} Harko, T. \& Lobo, F. S. N., \lq\lq{}Generalized Curvature-Matter Couplings in Modified Gravity,\rq\rq{} {\it Galaxies} {\bf 2} 410 (2014).
\bibitem{Zaregonbadi} Zaregonbadi, R., Farhoudi M., and Riazi N., \lq\lq{}Dark matter from $f({\sf R},{\sf T})$,\rq\rq{} {\it Phys. Rev. D} {\bf 94} 084052 (2016).
\bibitem{bertorotationc} Bertolami, O., Paramos, J., \lq\lq{}Mimicking dark matter through a non-minimal gravitational coupling with matter,\rq\rq{} JCAP 03 (2010) 009;\\ Bertolami, O., Paramos, J., \lq\lq{}Dark matter as a dynamic effect due to a non-minimal gravitational coupling with matter,\rq\rq{} J. Phys.: Conf. Ser. {\bf 222} (2010), 012010.
\bibitem{fRT2} Shabani, H., \& Farhoudi, M., \lq\lq{}$f({\sf R},{\sf T})$ cosmological models in phase-space\rq\rq{}, \textit{Phys. Rev. D} {\bf 88} (2013), 044048.
\bibitem{fRT3}     Kiani, F., \& Nozari, K., ``Energy conditions in $F({\sf T},\Theta)$ gravity and compatibility with a stable de                        Sitter solution", \textit{Phys. Lett. B} {\bf 728} (2014), 554.
\bibitem{fRT4}    Shabani, H., \& Farhoudi, M. ``cosmological and solar system consequences of $f({\sf R},{\sf T})$ gravity models",                    \textit{Phys. Rev. D} \textbf{88} (2014), 044031.
\bibitem{fRT5}   Azizi, T. \& Yaraie, E., ``G\"{o}del-type universes in Palatini $f({\sf R})$ gravity with a non-minimal                            curvature-matter coupling", \textit{Int. J. Theor. Phys.} {\bf 55} (2016), 176.
\bibitem{fRT6}    Alves, M. E. S., Moraes, P. H. R. S., de Araujo, J. C. N. \& Malheiro, M., \lq\lq{}Gravitational waves in $f({\sf R},{\sf T})$ and $f({\sf R};{\sf T}^{\phi})$ theories of gravity,\rq\rq{} \textit{Phys. Rev. D} \textbf{94} (2016), 024032.
\bibitem{fRT7}   Shabani, H. \& Ziaie, A. H.,``Stability of the Einstein static Universe in $f({\sf R},{\sf T})$ gravity",                             \textit{Eur. Phys. J.  C} {\bf 77} (2017), 31.
\bibitem{fRT8}   Poplawski, Nikodem J., ``A Lagrangian description of interacting dark energy", \textit{gr-qc/0608031}.
\bibitem{fRT9}    Ahmed, N. \& Pradhan, A.,\rq\rq{}Bianchi Type-V cosmology in f ({\sf R},{\sf T}) gravity with $\Lambda({\sf T})$\rq\rq{}, \textit{Int. J. Theor. Phys.} {\bf 53} (2014), 289.
\bibitem{fRT10} S.~D.~Odintsov and D.~Saez-Gomez,\lq\lq{}$f({\sf R}, {\sf T}, {\sf R}_{\mu\nu} {\sf T}^{\mu\nu})$ gravity phenomenology and $\Lambda$CDM Universe,\rq\rq{}  Phys.\ Lett.\ B {\bf 725}, 437 (2013);\\ Haghani, Z., Harko, T., Lobo, F. S. N., Sepangi, H. R. and Shahidi, S., \lq\lq{}Further matters in space-time geometry: $f({\sf R}, {\sf T}, {\sf R}_{\mu\nu} {\sf T}^{\mu\nu})$ gravity,\rq\rq{} {\it Phys. Rev. D} {\bf 88} (2013), 044023,
\bibitem{fRT11} F.~G.~Alvarenga, A.~de la Cruz-Dombriz, M.~J.~S.~Houndjo, M.~E.~Rodrigues and D.~Saez-Gomez,
\lq\lq{}Dynamics of scalar perturbations in f({\sf R},{\sf T}) gravity,\rq\rq{} Phys.\ Rev.\ D {\bf 87}, 103526 (2013)
;Erratum: [Phys.\ Rev.\ D {\bf 87}, no. 12, 129905 (2013)].
\bibitem{Baffou14} Baffou, E.H., Kpadonou, A.V., Rodrigues, M.E., Houndjo, M.J.S. \& Tossa, J., ``Cosmological viable $f (R,T)$ dark energy model: dynamics and stability", \textit{Astrophys. Space Sci.} {\bf355} (2014), 2197.
\bibitem{fRT12}    Shabani, H. \& Ziaie, A. H.\rq\rq{}Consequences of energy conservation violation: Late time solutions of $\Lambda$({\sf T})CDM subclass of f({\sf R},{\sf T}) gravity using dynamical system approach\rq\rq{}, \textit{Eur. Phys. J.  C} {\bf 77} (2017), 282.
\bibitem{fRT13} Shabani, H., \lq\lq{}Cosmological consequences and statefinder diagnosis of non-interacting generalized Chaplygin gas in f({\sf R},{\sf T}) gravity\rq\rq{}, \textit{Int. J. Mod. Phys. D.} {\bf 26} (2017), 1750120 .
\bibitem{fRT14} Harko, T., \lq\lq{}Thermodynamic interpretation of the generalized gravity models with geometry-matter coupling,\rq\rq{} Phys. Rev. D 90, 044067 (2014).
\bibitem{fRT15} Haghani, Z., Harko, T., Sepangi, H. R., Shahidi, S., \lq\lq{}Matter may matter,\rq\rq{} Int. J. Mod. Phys. D {\bf 23} (2014), 1442016.
\bibitem{fRT16} Xu, M.-X., Harko, T. and Liang, S.-D., \lq\lq{}Quantum Cosmology of $f({\sf R},{\sf T})$ gravity,\rq\rq{} {\it Eur. Phys. J. C}, {\bf 76} (2016), 1.
\bibitem{josset}  Josset, T., Perez, A., Sudarsky, D., \lq\lq{}Dark Energy from Violation of Energy Conservation\rq\rq{}, \textit{Phys. Rev. Lett.} {\bf 118} (2017), 021102.
\bibitem{violationcGR} A. S. AI-Rawaf I \& M. O. Taha, \lq\lq{}Cosmology of general relativity without energy-momentum conservation,\rq\rq{} Gen. Relativ. Gravit., {\bf 28} (1996) 935.
\bibitem{Jamil12} Jamil, M., Momeni, D., Raza, M. \& Myrzakulov, R., ``Reconstruction of some cosmological models in $f({\sf R},{\sf T})$ cosmology", \textit{Eur. Phys. J. C} {\b72} (2012), 1999.
\bibitem{Houndjo12} Houndjo, M.J.S.,``Reconstruction of $f({\sf R},{\sf T})$ gravity describing matter dominated and accelerated phases", \textit{Int. J. Mod. Phys. D} {\bf21} (2012), 1250003.
\bibitem{Singh14} Singh, C.P. \& Singh, V., ``Reconstruction of modified $f({\sf R},{\sf T})$ gravity with perfect fluid cosmological models", \textit{Gen. Relativ. Gravit.} {\bf46} (2014), 1696.
\bibitem{Sharif142} Sharif, M., Zubair, M., ``Reconstruction and stability of $f({\sf R},{\sf T})$ gravity with Ricci and modified Ricci dark energy", \textit{Astrophys. Space Sci.} {\bf349} (2014), 529.
\bibitem{Sharif141} Sharif, M., \& Zubair, M., ``Cosmological reconstruction and stability in $f({\sf R},{\sf T})$ gravity", \textit{Gen. Relativ. Gravit.} {\bf46} (2014), 1723.
\bibitem{Rudra15} Rudra, P., ``Does $f({\sf R},{\sf T})$  gravity admit a stationary scenario between dark energy and dark matter in its framework", \textit{Eur. Phys. J. Plus} {\bf 130} (2015), 66.
\bibitem{Baffou15}   Baffou, E.H., Houndjo, M.J.S., Rodrigues, M.E., Kpadonou, A.V.\& Tossa, J., ``Cosmological evolution in $f({\sf R},{\sf T})$ theory with collisional matter", \textit{Phys. Rev. D} {\bf 92} (2015), 084043.
\bibitem{Moraes161} Moraes, P.H.R.S., Ribeiro, G. \& Correa, R.A.C., ``A transition from a decelerated to an accelerated phase
of the Universe expansion from the simplest non-trivial polynomial function of T in the $f({\sf R},{\sf T})$ formalism", \textit{Astrophys. Space Sci.} {\bf361} (2016), 227.
\bibitem{SinghP14}   Singh, C.P., \& Kumar, P., ``Friedmann model with viscous cosmology in modified $f({\sf R},{\sf T})$
gravity theory", \textit{Eur. Phys. J.  C} {\bf 74} (2014), 3070.
\bibitem{Moraes15} Moraes, P.H.R.S., ``Cosmological solutions from induced matter model applied to 5D $f({\sf R},{\sf T})$ gravity and the shrinking of the extra coordinate", \textit{Eur. Phys. J.  C} {\bf 75} (2015), 168.
\bibitem{Yadav15} Yadav, Anil Kumar, Srivastava, P.K. \& Yadav Lallan, ``Hybrid Expansion Law for Dark Energy Dominated Universe in $f({\sf R},{\sf T})$ Gravity",  \textit{Int. J. Theor. Phys.} {\bf54} (2015), 1671.
\bibitem{Moraes162}  Moraes, P.H.R.S. \&  Correa, R.A.C., ``Evading the non-continuity equation in the $f({\sf R},{\sf T})$ formalism", arxiv:1606.07045 [gr-qc].
\bibitem{Sun16} Sun, G. \& Huang, Y.-C., ``The cosmology in $f({\sf R},{\sf T})$ gravity without dark energy", \textit{Int. J. Mod. Phys. D} {\bf25} (2016), 1650038.
\bibitem{Baffou171}  Baffou, E.H., Houndjo, M.J.S. \& Salako, I.G., ``Viscous generalized Chaplygin gas interacting with $f({\sf R},{\sf T})$ gravity", \textit{Int. J. Mod. Phys. D} {\bf14} (2017), 1750051.
\bibitem{Moraes163} Moraes, P.H.R.S., Correa, R.A.C.  \&  Ribeiro2, G., ``The Starobinsky model within the $f({\sf R},{\sf T})$ formalism as a cosmological model", arxiv:1701.01027 [gr-qc].
\bibitem{Baffou172}   Baffou, E.H., Houndjo, M.J.S., Hamani-Daouda, M. \& Alvarenga, F.G., ``Late time cosmological approach in mimetic $f({\sf R},{\sf T})$ gravity", arxiv:1706.08842 [gr-qc].
\bibitem{amend} Amendola, L., Gannouji, R., Polarski, D. \& Tsujikawa, S., ``Conditions for the cosmological viability of $f({\sf R})$ dark energy models'', \textit{Phys. Rev. D} \textbf{75} (2007), 083504.
\bibitem{berto-nonmini-cappo} Bertolami, O., Lobo, F. S. N., \& Paramos, J., \lq\lq{}Nonminimal coupling of perfect fluids to curvature\rq\rq{}, Phys. Rev. D {\bf 78} (2008), 064036;\\ Bertolami, O., Harko, T., Lobo, F. S. N. \& Paramos, J., \lq\lq{}Non-minimal curvature-matter couplings in modified gravity\rq\rq{}, arXiv:0811.2876 [gr-qc];\\ Harko, T., Lobo, F. S. N., \lq\lq{}$f({\sf R},{\sf L}_m)$ gravity\rq\rq{}, Eur. Phys. J. C {\bf 70} (2010), 373;\\ Ludwig, H., Minazzoli O. \& Capozziello, S., \lq\lq{}Merging matter and geometry in the same Lagrangian\rq\rq{}, Phys. Lett. B {\bf 751} (2015), 576.
\bibitem{farhoudipps} Farhoudi, M. \lq\lq{}Classical trace anomaly\rq\rq{}, Int. J. Mod. Phys. D {\bf 14} (2005), 1233;\\ Farhoudi, M., \lq\lq{}Non-linear Lagrangian Theories of Gravitation\rq\rq{}, (Ph.D. thesis, Queen Mary and Westfield College, University of London, 1995).
\bibitem{Kharko} Harko, T., \lq\lq{}Modified gravity with arbitrary coupling between matter and geometry\rq\rq{}, Phys. Lett. B {\bf 669} (2008), 376.
\bibitem{Brown-HAwking} Brown, J. D. Class. Quantum Grav. {\bf 10} (1993) 1579.
\bibitem{Brown-HAwking1} S. W. Hawking and G. F. R. Ellis, \lq\lq{}{The Large Scale Structure of Spacetime}\rq\rq{} (Cambridge University Press, Cambridge, 1973)

\bibitem{Hwang} Hwang, J. C., \lq\lq{}Perturbations of the Robertson-Walker space: multicomponent sources and generalized gravity\rq\rq{}, \textit{Astrophys. J} \textbf{375} (1991), 443.
\bibitem{Amend} Amendola, L., Gannouji, R., Polarski, D. \& Tsujikawa, S., \lq\lq{}Conditions for the cosmological viability of $f({\sf R})$ dark energy models\rq\rq{}, \textit{Phys. Rev. D} \textbf{75} (2007), 083504.
\bibitem{Fay} Fay, S., Nesseris, S. \& Perivolaropoulos, L., \lq\lq{}Can $f({\sf R})$ modified gravity theories mimic a $\Lambda CDM$ cosmology?\rq\rq{}, \textit{Phys. Rev. D} \textbf{76} (2007), 063504.
\bibitem{Elizalde} Elizalde, E., Odintsov, S.D., Sebastiani, L. \& Zerbini, S., \lq\lq{}Oscillations of the F({\sf R}) dark energy in the accelerating Universe\rq\rq{}, \textit{Eur. Phys. J. C} \textbf{72} (2012), 1843.
\bibitem{Oikonomou} Oikonomou, V. K. , \lq\lq{}An exponential $F({\sf R})$ dark energy model\rq\rq{}, \textit{Gen. Relativ. Gravit.} \textbf{45} (2013), 2467.
\bibitem{Mukherjee} Mukherjee, A. \& Banerjee, N., \lq\lq{}Acceleration of the Universe in $f({\sf R})$ gravity models\rq\rq{}, \textit{Astrophys. Space Sci.} \textbf{352} (2014), 893.
\bibitem{Bamba} Bamba,K. \& Odintsov,S.D., \lq\lq{}Universe acceleration in modified gravities: $ F({\sf R}) $ and $ F ({\sf T}) $ cases\rq\rq{}, \textit{hep-th/1402.7114}
\bibitem{Cosmai} Cosmai, L., Fanizza, G. \& Tedesco, L.,``Cosmic Acceleration and $f ({\sf R})$ Theory: Perturbed Solution in a Matter FLRW Model", \textit{Int. J. Theor. Phys.} \textbf{55} (2016), 754.
\bibitem{Battye} Battye, Richard A., Bolliet, Boris, Pearson \& Jonathan A.,\rq\rq{}$f({\sf R})$ gravity as a dark energy fluid\rq\rq{}', \textit{Phys. Rev. D} \textbf{93} (2016), 044026.
\bibitem{Shabani6} Shabani, H. \& Ziaie, A. H., ``Interpretation of $f({\sf R},{\sf T})$ gravity in terms of a conserved effective fluid ", arXiv:1702.07380 [gr-qc]
\bibitem{Prigogine} Prigogine, I., Geheniau, J., Gunzig, E. \& Nardone, P. ``Thermodynamics of cosmological matter creation",\textit{Proc. Nati. Acad. Sci.} \textbf{85} (1988),7428. 
\end{thebibliography}
\end{document}